\documentclass[titlepage,11pt]{article}%
\usepackage{graphicx}
\usepackage{amsmath}
\usepackage{amsfonts}
\usepackage{amssymb}%
\begin{document}

\title{{\LARGE Annuitization and Asset Allocation}}
\author{Moshe A. Milevsky\thanks{Milevsky, the contact author, is an Associate
Professor of Finance at the Schulich School of Business, York University,
Toronto, Ontario, M3J 1P3, Canada, and the Director of the Individual Finance
and Insurance Decisions (IFID) Centre at the Fields Institute. He can be
reached at Tel: (416) 736-2100 ext 66014, Fax: (416) 763-5487, E-mail:
milevsky@yorku.ca. This research is partially supported by a grant from the
Social Sciences and Humanities Research Council of Canada and the Society of
Actuaries. \medskip} and Virginia R. Young\thanks{Young is a Professor of
Mathematics at the University of Michigan, Ann Arbor, Michigan, 48109, USA.
She can be reached at Tel: (734) 764-7227, Fax: (734) 764-7048, E-mail:
vryoung@umich.edu. The authors acknowledge helpful comments from two anonymous
JEDC referees, Jeff Brown, Narat Charupat, Glenn Daily, Jerry Green, Mike
Orszag, Mark Warshawsky, and seminar participants at \emph{York University},
\emph{University of Michigan, University of Wisconsin}, \emph{University of
Cyprus}, the \emph{American Economics Association} and the \emph{Western
Finance Association }annual meeting. }\\York University and University of Michigan}
\date{Current Version: 25 August 2006}
\maketitle

\begin{abstract}
\ \ \ \ \ \ \ \ \ \ \ \ \ \ \ \ \ \ \ \ \ \ \ \ \ \ \ \ \ \textbf{ANNUITIZATION
AND ASSET ALLOCATION}

This paper examines the optimal annuitization, investment and consumption
strategies of a utility-maximizing \emph{retiree} facing a stochastic time of
death under a variety of institutional restrictions. We focus on the impact of
aging on the optimal purchase of life annuities which form the basis of most
Defined Benefit pension plans. Due to adverse selection, acquiring a lifetime
payout annuity is an irreversible transaction that creates an incentive to
delay. Under the institutional all-or-nothing arrangement where annuitization
must take place at one distinct point in time (i.e. retirement), we derive the
optimal age at which to annuitize and develop a metric to capture the loss
from annuitizing prematurely. In contrast, under an open-market structure
where individuals can annuitize any fraction of their wealth at anytime, we
locate a general optimal annuity purchasing policy. In this case, we find that
an individual will initially annuitize a lump sum and then buy annuities to
keep wealth to one side of a separating \emph{ray} in wealth-annuity space. We
believe our paper is the first to integrate life annuity products into the
portfolio choice literature while taking into account realistic institutional
restrictions which are unique to the market for mortality-contingent claims.

\medskip


\medskip


\end{abstract}

%

%TCIMACRO{\TeXButton{TeX field}{\newpage}}%
%BeginExpansion
\newpage
%EndExpansion

\section{Introduction and Motivation}

Asset allocation and consumption decisions towards the end of the human life
cycle are complicated by the uncertainty associated with the length of life.
Although this risk can be completely hedged in a \emph{perfect market} with
life annuities -- or, more precisely, with continuously-renegotiated tontines
-- real world frictions and imperfections impede the ability to do so in
practice. Indeed, empirical and anecdotal evidence suggests that voluntary
annuitization amongst the public is not very common, nor is it well understood
even amongst financial advisors. Therefore, in attempt to fill this gap and
integrate mortality-contingent claims into the finance literature, this paper
examines the optimal annuitization strategy of a utility-maximizing
\emph{retiree} facing a stochastic time of death under a variety of
institutional pension and annuity arrangements. We also examine the usual
investment and consumption dynamics but focus our attention on the impact of
aging and the increase in the actuarial force of mortality on the optimal
purchase of mortality-contingent annuities, which form the basis of most
Defined Benefit (DB) pension plans. One of our main insights is that due to
severe adverse selection concerns, acquiring a lifetime payout annuity is an
irreversible transaction that, we argue, creates an incentive to delay. We
appeal to the analogy of a classical American option which should only be
exercised once the value from waiting is no more than the value from exercising.

For the most part of the paper, the focus of our attention is a life annuity
that pays a fixed (real or nominal) continuous payout for the duration of the
annuitant's life. (The appendix extends our basic model to variable immediate
annuities.) From a financial perspective, this product is akin to a
coupon-bearing bond that defaults upon death of the holder and for which there
is no secondary market. Under the institutional all-or-nothing arrangement,
where annuitization (i.e. the purchase) must take place at one distinct point
in time (i.e. retirement), we locate the optimal age at which to annuitize and
develop a metric to capture the loss from annuitizing prematurely. This
optimal age, which is linked to the actuarial force of mortality, occurs
within retirement years and is obviously gender specific but also depends on
the individual's subjective health status. All of this will be explained in
the body of the paper.

In contrast to the restrictive (yet not uncommon) all-or-nothing arrangement,
under an open-market structure where individuals can annuitize a fraction of
their wealth at distinct points in time, we locate a general optimal annuity
purchasing policy. In this case, we find that an individual will initially
annuitize a lump sum -- if they do not already have this minimum level in
pre-existing DB pensions -- and then buy additional life annuities in order to
keep wealth to one side of a separating \emph{ray} in wealth-annuity space.
This is a type of barrier control result that is common in the literature on
asset allocation with transaction costs.

We believe our paper is the first\footnote{The first (working paper) version
of this paper was circulated and presented at the January 2001 meeting of the
\textit{American Economics Association} in New Orleans.} to integrate life and
pension annuity products into the portfolio choice literature while taking
into account realistic institutional restrictions which are unique to the
market for mortality-contingent claims.

\subsection{Agenda and Outline}

The remainder of this paper is organized as follows. In Section 2, we provide
a brief explanation of the mechanics of the life annuity market and review the
existing literature involving asset allocation, personal pensions, and payout
annuities. In Section 3, we present the general model for our financial and
annuity markets. In Section 4, we consider the case for which the individual
is required to annuitize \emph{all} her pensionable wealth at one point in
time. This is effectively an optimal retirement problem and is akin to the
situation (up until recently) in the United Kingdom, where retirees can
drawdown their pension but must annuitize the remaining balance by a certain
age, or to the situation for which individuals have the choice of when to
start their retirement (DB) pension but must do so at \emph{one} point in
time. In fact, most Variable Annuity contracts sold in the United States have
an embedded option to annuitize that can only be exercised once. Our analysis
would cover this too. In this restrictive (but common) framework, we locate
the optimal age for her to do so, and then define a so-called option value
\emph{metric} as the gain in utility from annuitizing optimally. Section 5
provides a variety of numerical examples for the optimal time to annuitize and
also pursues the option analogy as a way of illustrating the loss from
annuitizing pre-maturely.

Then, in Section 6, we consider a less restrictive open-market arrangement
whereby the individual may annuitize any portion of her wealth at any time.
This arrangement is applicable to individuals with substantial discretionary
wealth who can purchase small (or large) quantities of annuities on an ongoing
basis. In this case, we find that the individual annuitizes a lump sum as soon
as possible (the amount might be zero) and then acquires more annuities
depending on the performance of her stochastic wealth process. If her wealth
subsequently increases in value, she purchases more annuities by annuitizing
additional wealth; otherwise, she refrains from additional purchases and
consumes from her originally-purchased annuities, as well as from liquidating
investments in her portfolio. Furthermore, we \emph{explicitly} solve for the
optimal annuity purchasing policy under this less restrictive case when the
force of mortality is constant, which implies that the future lifetime is
exponentially distributed. Section 7 provides a variety of numerical examples
for the open-market framework and flushes-out and explores a number of
insights. Non-essential proofs and theorems are relegated to an Appendix
(Section 9), while Section 8 concludes the paper with our main qualitative insights.

\section{The Annuity Market and Literature}

\subsection{Life Annuity Basics}

Life annuities are purchased directly from insurance companies and form the
basis of most DB pension plans. In exchange for a lump-sum premium, which the
company invests in its general account, the company guarantees to pay the
annuitant a fixed (monthly or quarterly) payout for the rest of his or her
life. This payout rate -- which depends on prevailing interest rates and
mortality projections -- is irrevocably determined at the time of purchase
(a.k.a. annuitization) and does not change for the life of the contract. The
following chart illustrates some sample quotes which were provided by a
life-annuity broker (in Canada).

\begin{center}
$%
\begin{tabular}
[c]{|c|c|c|c|c|c|c|c|c|c|c|c|c|}\hline
Certain
%TCIMACRO{\TEXTsymbol{\backslash} }%
%BeginExpansion
$\backslash$
%EndExpansion
Age & \multicolumn{2}{|c|}{m \textbf{55 }f} & \multicolumn{2}{|c|}{m
\textbf{60 }f} & \multicolumn{2}{|c|}{m \textbf{65 }f} &
\multicolumn{2}{|c|}{m \textbf{70 }f} & \multicolumn{2}{|c|}{m \textbf{75 }f}
& \multicolumn{2}{|c|}{m \textbf{80 }f}\\\hline
\textbf{0 yrs} & 631 & 590 & 686 & 633 & 765 & 694 & 877 & 780 & 1039 & 911 &
1259 & 1096\\\hline
\textbf{5 yrs} & 628 & 589 & 681 & 631 & 755 & 689 & 855 & 770 & 989 & 887 &
1146 & 1036\\\hline
\textbf{10 yrs} & 620 & 584 & 666 & 623 & 726 & 674 & 799 & 741 & 879 & 825 &
940 & 901\\\hline
\textbf{15 yrs} & 607 & 578 & 644 & 611 & 687 & 652 & 729 & 699 & 764 & 745 &
774 & 765\\\hline
\textbf{20 yrs} & 591 & 569 & 618 & 596 & 643 & 625 & 662 & 651 & 673 & 668 &
665 & 664\\\hline
\textbf{25 yrs} & 573 & 559 & 589 & 578 & 601 & 594 & 608 & 605 & 610 & 609 &
N.A. & N.A.\\\hline
\multicolumn{13}{|c|}{$\text{{\small Sample monthly payout based on a
\$100,000 initial premium (purchase)}}$}\\\hline
\end{tabular}
\ $
\end{center}

The term word certain refers to the guarantee period built into the annuity.
If the guarantee period is $n$ years, then the individual buying such an
annuity (or his or her estate) will receive the stated income for $n$ years;
thereafter, the individual will receive the money \textit{only} if he or she
is alive.

For example, in exchange for a \$100,000 initial premium, a 75-year-old female
will receive \$911 per month for the rest of her life. This life annuity has
no guarantee period, which means that if she were to die one instant after
purchasing the life annuity (technically it would have to be after the first
payment), her beneficiaries or estate would receive nothing in return. The
\$911 monthly income for those who survive consists of a mix of principal and
interest as well as the implicit funds of those who do not survive. A male
would receive slightly more per month, namely \$1,039, due to the lower life
expectancy of males.

A few things should be obvious from the table. First, the higher the purchase
age, all else being equal, the greater the annuity income. In this case, the
future life expectancy is shorter and the initial premium must be amortized
and returned over a shorter time period. Likewise, a longer guarantee period
yields a lower annuity income. In fact, an 80-year-old female buying a 20-year
guarantee will receive virtually the same amount (\$664) as a male of the same
age, since neither is likely to live past the 20-year certain period; hence,
the annuity is essentially a portfolio of zero-coupon bonds. Some other points
are in order, especially for those not familiar with insurance pricing concepts.

\begin{enumerate}
\item The law of large numbers and the ability to diversify mortality risk is
central to the pricing of life annuities. The above-mentioned payouts are
determined by expected \emph{objective} annuitant mortality patterns together
with prevailing interest rates of corresponding durations. Profits, fees, and
commissions are built into these quotes by \emph{loading} the pure actuarial
factor on the order of 1\% to 5\%.

\item Payout rates fluctuate from week-to-week because most of the insurance
companies' assets backing these lifetime guarantees are invested in
fixed-income instruments. This can sometimes cause quotes to change on a daily
basis. It is therefore reasonble to model the evolution of these prices in
continuous time.

\item Most of the existing open-purchase (a.k.a retail) annuity market in
North America is based on fixed nominal (and not inflation-adjusted) payouts.
Real annuities are quite rare, which is an ongoing puzzle to many economists.
Consequently, most of the numerical examples in our paper focus on nominal
values, although there is nothing in our model that precludes using
inflation-adjusted prices and returns as long as they are not mixed in the
same model.

\item An additional form of life annuity is the variable payout kind whose
periodic income is linked to the performance of pre-selected equity and bond
indices. In this case, the above-mentioned \$100,000 premium would go towards
purchasing a number of payout \emph{units} (as opposed to dollars) whose value
would fluctuate over time. These annuities are the foundation of the US-based
TIAA-CREF's pension plan for University workers, but are quite rare anywhere
else in the world. As a result, the bulk of our paper addresses the fixed
payout kind, but we refer the interested reader to Appendix C (namely, Section
9.3) in which these products are integrated into our model.
\end{enumerate}

\subsection{Literature Review}

This paper merges a variety of distinct strands in the portfolio choice and
annuity literature. First, our work sits squarely within the classical Merton
(1971) optimal asset allocation and consumption framework. However, in
contrast to extensions of this model by Kim and Omberg (1996), Koo (1998),
Sorensen (1999), Wachter (2002), Bodie, Detemple, Otruba, and Walter (2004),
or the recent book by Campbell and Viceira (2002), for example -- which are
concerned with relaxing the dynamics of the underlying state variables and/or
investigating the impact of (retirement) time horizon on portfolio choice --
our model attempts to realistically incorporate mortality-contingent payout
annuities within this framework.

A life-contingent annuity is the building block of most DB pension plans --
see Bodie, Marcus, and Merton (1988) for details -- but can also be purchased
in the retail market. The irreversibility of this purchase is due to the
well-known adverse selection issues identified by Akerlof (1970) and
Rothschild and Stiglitz (1976).

On its own, the topic of payout annuities has been investigated quite
extensively within the public economics literature. In fact, a so-called
annuity puzzle has been identified in this field. The puzzle relates to the
incredibly low levels of voluntary annuitization exhibited by retirees who are
given the choice of purchasing a mortality-contingent payout annuity. For
example, holders of \emph{variable annuity saving} policies in the U.S. have
the option to convert their accumulated savings into a payout annuity, and yet
less than 2\% elect to do so according to the National Association of Variable
Annuities and LIMRA International (www.limra.com). In the comprehensive Health
and Retirement Survey (HRS) conducted in the U.S, only 1.57\% of the HRS
respondents reported life annuity income. Likewise, only 8.0\% of respondents
with a defined contribution pension plan selected an annuity payout.

Collectively, these low levels of voluntary annuitization stand in contrast to
the implications of the Modigliani life-cycle hypothesis, as pointed out in
Modigliani's (1986) Nobel prize lecture. Indeed, as originally demonstrated by
Yaari (1965), individuals with no utility of bequest should hold all their
assets in mortality-contingent annuities since they stochastically dominate
the payout from conventional asset classes. The result of Yaari (1965) has
been the subject of much research in the public economics and insurance
literature, and we refer the interested reader to a series of papers by
Friedman and Warshawsky (1990), Brown (1999, 2001), Mitchell, Poterba,
Warshawsky, and Brown (MPWB) (1999), Brown and Poterba (2000), and Brown and
Warshawsky (2001). Collectively, these papers place some of the `blame' for
low annuitization rates on the high loads and fees that are embedded in
annuity prices. Other economic-based explanations include Kotlikoff and
Summers (1981), Kotlikoff and Spivak (1981), Hurd (1989), and Bernhiem (1991),
which focus on the role of families and their bequest motives on the demand
for annuitization. Other models that focus on market imperfections and adverse
selection include Brugiavini (1993) and Yagi and Nishigaki (1993).

Thus, given the rich literature on dynamic asset allocation and the increasing
interest in pension-related finance issues, our objective is to incorporate
longevity-insurance products into a portfolio and asset allocation framework
that properly captures the actuarial and insurance imperfections. Although
Richard (1975) extended Merton's (1971) model to obtain Yaari's (1965) results
in a continuous-time framework, the institutional set-up lacked the realism of
current payout annuity markets.\footnote{Recent papers that attempt a
portfolio-based model for annuitization along the same lines -- most written
after the first draft of this paper was released -- include Kapur and Orszag
(1999), Blake, Cairns, and Dowd (2000), Cairns, Blake, and Dowd (2005),
Neuberger (2003), Dushi and Webb (2003), Sinclair (2003), Stabile (2003), and
Battocchio, Menoncin, and Scaillet (2003), Koijen, Nijman and Werker (2006).}

We also argue that the decision of when to purchase an irreversible life
annuity endows the holder with an incentive to delay that can be heuristically
viewed as an option. Indeed, under many institutional pension arrangements
(such as in the United Kingdom up until recently or with regards to the rules
concerning variable annuity saving policies in the U.S.), individuals are
allowed to drawdown their pension via discretionary consumption but must
eventually annuitize at \emph{one point in time} their remaining wealth. We
refer to this system as an \underline{all-or-nothing} arrangement and argue
that this is similar to Stock and Wise's (1990) option to retire and echoes
the framework of Sundaresan and Zapatero (1997) who examine optimal behavior
(and valuation) of various pension benefits. Other institutional structures
allow for annuitization at any time and in small quantities as well, and we
refer to these systems as \underline{anything anytime} throughout the paper.
We investigate the optimal annuitization policy in both of these cases and
provide extensive numerical examples that compare the two.

\section{Financial and Pension Annuity Markets}

In this section, we describe our model for financial and annuity markets. We
assume that an individual can invest in a riskless asset whose price at time
$s$, $X_{s}$, follows the process $dX_{s}=rX_{s}ds,X_{t}=X>0$, for some fixed
$r\geq0$. Also, the individual can invest in a risky asset whose price at time
$s$, $S_{s}$, follows geometric Brownian motion given by%

\begin{equation}
\left\{
\begin{array}
[c]{l}%
dS_{s}=\mu S_{s}ds+\sigma S_{s}dB_{s}\text{,}\\
S_{t}=S>0\text{,}%
\end{array}
\right.
\end{equation}

\noindent in which $\mu>r,$ $\sigma>0$, and $B$ is a standard Brownian motion
with respect to a filtration $\{\mathcal{F}_{s}\}$ of the probability space
$(\Omega, \mathcal{F}, \mathbf{P})$. Let $W_{s}$ be the wealth at time $s$ of
the individual, and let $\pi_{s}$ be the amount that the decision maker
invests in the risky asset at time $s$. Also, the decision maker consumes at a
rate of $c_{s}$ at time $s$. Then, the amount in the riskless asset is
$W_{s}-\pi_{s}$, and when the individual buys no annuities, wealth follows the process%

\begin{equation}
\left\{
\begin{array}
[c]{rl}%
dW_{s} & =d\left(  W_{s}-\pi_{s}\right)  +d\pi_{s} - c_{s}ds\\
& =r\left(  W_{s}-\pi_{s}\right)  dt+\pi_{s}\left(  \mu ds+\sigma
dB_{s}\right)  -c_{s}ds\\
& =\left[  rW_{s}+\left(  \mu-r\right)  \pi_{s}-c_{s}\right]  ds+\sigma\pi
_{s}dB_{s}\text{,}\\
W_{t} & =w>0\text{.}%
\end{array}
\right.  \label{2.1}%
\end{equation}

In Sections 4 and 5, we assume that the decision maker seeks to maximize (over
admissible $\{c_{s},\pi_{s}\}$ and over times of annuitizing all his or her
wealth, $\tau)$ the expected utility of discounted consumption. Admissible
$\{c_{s},\pi_{s}\}$ are those that are measurable with respect to the
information available at time $s$, namely $\mathcal{F}_{s}$, that restrict
consumption to be non-negative, and that result in (\ref{2.1}) having a unique
solution; see Karatzas and Shreve (1998). We also allow the individual to
value expected utility via a subjective hazard rate (or force of mortality),
while the annuity is priced by using an objective hazard rate; which may or
may not be the same.

Our financial economy is based on the (simpler) geometric Brownian motion plus
risk-free rate model originally pioneered by Merton (1971), as opposed to the
more recent and richer models developed by Kim and Omberg (1996), Sorensen
(1999), Wachter (2002), or Campbell and Viceira (2002) for example. The reason
is that we are primarily interested in the implications of introducing a
mortality-contingent claim into the portfolio choice framework, as opposed to
studying the impact of stochastic interest rates or mean-reverting equity
premiums \textit{per se}. By avoiding the computational price of a more
complex set-up, we are able to obtain analytical solutions to our
annuitization problems.

We now move on to the insurance and actuarial assumptions. We let $_{t}
p_{x}^{S}$ denote the subjective conditional probability that an individual
aged $x$ believes he or she will survive to age $x+t$. It is defined via the
subjective hazard function, $\lambda_{x+t}^{S}$, by the formula%

\begin{equation}
_{t}p_{x}^{S}=\exp\left(  -\int_{0}^{t}\lambda_{x+s}^{S} \; ds\right)  .
\end{equation}

\noindent We have a similar formula for the objective conditional probability
of survival, $_{t}p_{x}^{O},$ in terms of the objective hazard function,
$\lambda_{x+t}^{O}$.

The actuarial present value of a life annuity that pays \$1 per year
continuously to an individual who is age $x$ at the time of purchase is
written $\bar{a}_{x}$. It is defined by%

\begin{equation}
\bar{a}_{x}=\int_{0}^{\infty}e^{-rt} \; {_{t}p_{x}} \; dt .
\end{equation}

\noindent We deliberately use the risk-free rate $r$ in our annuity pricing
because most of the recent empirical evidence suggests that the \emph{money's
worth} of annuities relative to the risk-free Government yield curve is
relatively close to one. In other words, the expected present value of payouts
using the risk-free rate \emph{is} equal to the premium paid for that benefit.
Thus, it appears that the additional credit risk that the insurance company
might take on by investing in higher risk bonds is offset by any insurance
loads and commissions they charge. We refer the interested reader to the paper
by MPWB (1999) for a greater discussion of the precise curve that is used for
pricing in practice.

In terms of notation, if we use the subjective hazard rate to calculate the
survival probabilities in equation (4), then we write $\bar{a}_{x}^{S},$ while
if we use the objective (pricing) hazard rate to calculate the survival
probabilities, then we write $\bar{a}_{x}^{O}.$ Just to clarify, by objective
$\bar{a}_{x}^{O}$, we mean the actual market prices of the annuity net of any
insurance loading, whereas $\bar{a}_{x}^{S}$ denotes what the market price
`would have been' had the insurance company used the individual's personal and
subjective assessment of her mortality.

We refer the interested reader to Hurd and McGarry (1995, 1997) for a
discussion of experiments involving \textquotedblleft
subjective\textquotedblright\ versus \textquotedblleft
objective\textquotedblright\ assessments of survival
probabilities.\footnote{Also, more recently, Smith, Taylor, and Sloan (2001)
claim that their \textquotedblleft findings leave little doubt that subjective
perceptions of mortality should be taken seriously.\textquotedblright\ They
state that individuals' \textquotedblleft longevity expectations are
reasonably good predictions of future mortality.\textquotedblright\ Other
researchers, such as Bhattacharya, Goldman and Sood (2003) claim that
individuals are biased in their estimates of mortality as evidenced by the
viatical and life settlement market. We do not take a position on whether
individuals estimate and discount mortality using the same forward curve as
the insurance company and therefore allow for the two functions to be
distinct.} We will demonstrate that asymmetry of mortality beliefs might go a
long way towards explaining why individuals who believe themselves to be less
healthy than average are more likely to avoid annuities, despite having no
declared bequest motive. In the classical \emph{perfect market} Yaari (1965)
framework, subjective survival rates do not play a role in the optimal policy.
We will show that if the consumer disagrees with the insurance company's
pricing basis regarding her subjective hazard rate - or personal health status
- she will delay annuitization in an all-or-nothing environment.

In Section 6, we start by assuming that the decision maker maximizes (over
admissible $\{c_{s},\pi_{s},A_{s}\}$) the expected utility of discounted
lifetime consumption as well as bequest, in which $A_{s}$ is the annuity
purchasing process. $A_{s}$ denotes the non-negative annuity income
\textit{rate} at time $s$ after any annuity purchases at that time; we assume
that $A_{s}$ is right-continuous with left limits. The source of this income
could be previous annuity purchases or a pre-existing annuity, such as Social
Security or pension income. We assume that the individual can purchase an
annuity at the price of $\bar{a}_{x+s}^{O}$ per dollar of annuity income at
time $s$, or equivalently, at age $x+s$. In that case, the dynamics of the
wealth process are given by%

\begin{equation}
\left\{
\begin{array}
[c]{l}%
dW_{s}=\left[  rW_{s-}+\left(  \mu-r\right)  \pi_{s}-c_{s} + A_{s-} \right]
ds+\sigma\pi_{s} \, dB_{s} - \bar{a}_{x+s}^{O} \, dA_{s}\text{,}\\
W_{t-}=w>0\text{.}%
\end{array}
\right.  \label{2.2}%
\end{equation}

\noindent The negative sign on the subscripts for wealth and annuities denotes
the left-hand limit of those quantities before any (lump-sum) annuity purchases.

\section{Restricted Market: All or Nothing}

In this section, we examine the institutional arrangement where the individual
is required to annuitize all her wealth in a lump sum at some (retirement)
time $\tau$. If the volatility of investment return $\sigma=0$, and assuming
the objective hazard rate increases over time, then we show that the
individual annuitizes \textit{all} her wealth at a time $T$ for which
$\mu=r+\lambda_{x+T}^{O}$, which is the time at which the hazard rate (a.k.a.
mortality credits) plus the risk-free rate is equal to the expected return
from the asset. Furthermore, if $\lambda_{x+t}^{S}=\lambda_{x+t}^{O}$ for all
$t>0$, then the individual will optimally consume exactly the annuity income
after time $T$. Therefore, for small values of $\sigma$ and for $\lambda
_{x+t}^{S}\approx\lambda_{x+t}^{O}$ for all $t\geq0$, the individual will
consume \textit{approximately} the annuity income after annuitizing her
wealth, at least for time soon after the annuitization time. In fact, the
classical annuity results, for example Yaari (1965), prove that consuming the
entire annuity income is optimal in the absence of bequest motives. This is
why, to simplify our work we assume that at some time $\tau$, the individual
annuitizes all her wealth $W_{\tau}$ and thereafter consumes at a rate of
$\frac{W_{\tau}}{\bar{a}_{x+\tau}^{O}}$, the annuity income. 

It follows that the associated value function of this problem is given by%

\begin{align}
&  U(w,t)\nonumber\\
&  =\sup_{\left\{  c_{s},\pi_{s},\tau\right\}  }\mathbf{E}^{w,t}\left[
\int\nolimits_{t}^{\tau}e^{-r\left(  s-t\right)  }\left.  _{s-t}\right.
p_{x+t}^{S}\;u\left(  c_{s}\right)  ds+\int_{\tau}^{\infty}e^{-r\left(
s-t\right)  }\left.  _{s-t}\right.  p_{x+t}^{S}\;u\left(  \frac{W_{\tau}}%
{\bar{a}_{x+\tau}^{O}}\right)  ds\right] \nonumber\\
&  =\sup_{\left\{  c_{s},\pi_{s},\tau\right\}  }\mathbf{E}^{w,t}\left[
\int_{t}^{\tau}e^{-r\left(  s-t\right)  }\left.  _{s-t}\right.  p_{x+t}%
^{S}\;u\left(  c_{s}\right)  ds+e^{-r\left(  \tau-t\right)  }\left.  _{\tau
-t}\right.  p_{x+t}^{S}\;u\left(  \frac{W_{\tau}}{\bar{a}_{x+\tau}^{O}%
}\right)  \bar{a}_{x+\tau}^{S}\right]  , \label{3.1}%
\end{align}

\noindent in which $u$ is an increasing, concave utility function of
consumption, and $\mathbf{E}^{w,t}$ denotes the expectation conditional on
$W_{t}=w$. Note that the individual discounts consumption at the riskless rate
$r$. If we were to model with a subjective discount rate of say $\rho$, then
this is equivalent to using $r$ as in (\ref{3.1}) and adding $\rho-r$ to the
subjective hazard rate. Thus, there is no effective loss of generality in
setting the subjective discount rate equal to the riskless rate $r$. In other
words, while some life-cycle models in the literature adjust the discount rate
for perceived risk and other subjective factors, we remind the reader that our
underlying hazard rate $\lambda_{x+t}^{S}$ effectively adjusts the discount
rate for the probability of survival and, thus, takes these risks into account implicitly.

Note that in this section, we do not account for pre-existing annuities, such
as state and corporate pensions for example. We anticipate that such annuities
will change the optimal time of annuitization, but we defer this problem to
Section 6. Also, note that we take the annuity prices as exogenously given. We
are not creating an equilibrium (positive) model of pricing as in the adverse
selection literature of Akerlof (1970) or Rothschild and Stiglitz (1976), but
rather a normative model of how people should behave in the presence of these
given market prices. Expanding to equilibrium considerations is beyond the
scope of this (normative) paper, although we do make some statements regarding
equilibrium pricing of annuities in the concluding remarks.

We also restrict our attention to the case in which the utility function
exhibits constant relative risk aversion (CRRA), $\gamma=-cu^{\prime\prime
}\left(  c\right)  /u^{\prime}\left(  c\right)  $. That is, $u$ is given by%

\begin{equation}
u\left(  c\right)  =\frac{c^{1-\gamma}}{1-\gamma}, \quad\gamma>0,\gamma
\not =1. \label{utility}%
\end{equation}

\noindent For this utility function, the relative risk aversion equals
$\gamma$, a constant. The utility function that corresponds to relative risk
aversion 1 is logarithmic utility.

We next show that for CRRA utility, solving the problem in (6) is equivalent
to assuming that the optimal stopping annuitization time is some fixed time in
the future, say $T$. Based on that value of $T$, one finds the optimal
consumption and investment policies. Finally, one finds the optimal value of
$T \ge0$.

The feature of CRRA utility that drives this result is that wealth factors out
of the value function; therefore, the stopping time $\tau$ is not random, but
rather deterministic. For other utility functions, $\tau$ will be a random
stopping time that depends on stochastic (state variable of) wealth, as is the
case for the exercise time of American options.

To show that the optimal stopping time is deterministic for CRRA utility, we
use the fact that if we can find a smooth solution $V$ to the following
variational inequality, then that smooth solution equals the value function
$U$ in (\ref{3.1}); see \O ksendal (1998, Chapter 10), for example.%

\begin{equation}
\left(  r+\lambda_{x+t}^{S}\right)  V\geq V_{t}+rwV_{w}+\max_{c}\left[  u(c)
-cV_{w}\right]  +\max_{\pi}\left[  (\mu- r) \pi V_{w}+\frac{1}{2}\sigma^{2}%
\pi^{2}V_{ww}\right]  , \label{8.1}%
\end{equation}

\noindent and%

\begin{equation}
V(w, t) \ge\bar{a}_{x+t}^{S} \, u \left(  \frac{w}{\bar{a}_{x+t}^{O}}\right)
, \label{8.2}%
\end{equation}

\noindent with equality in at least one of (\ref{8.1}) and (\ref{8.2}), and
with $u$ given by (\ref{utility}).

We look for a solution of this variational inequality of the form $V(w, t)
=\frac{1}{1-\gamma}w^{1-\gamma} \psi^{\gamma}(t)$. If $V$ is of this form,
then $\psi$ necessarily solves

\bigskip%
\begin{equation}
\frac{1}{1-\gamma}\left(  r+\lambda_{x+t}^{S}\right)  \psi\geq\frac{\gamma}{1-
\gamma} \psi^{\prime} + \delta\psi+ \frac{\gamma}{1-\gamma}, \label{8.3}%
\end{equation}

\noindent and%

\begin{equation}
\frac{1}{1-\gamma}\psi^{\gamma}(t) \ge\frac{1}{1-\gamma}\frac{\bar{a}%
_{x+t}^{S}}{\left(  \bar{a}_{x+t}^{O}\right)  ^{1-\gamma}}, \label{8.4}%
\end{equation}

\noindent with equality in at least one of (\ref{8.3}) and (\ref{8.4}). So, if
we find a smooth solution $\psi$ to this variational inequality in time $t$,
then we are done, and we can assert that $U(w, t) = \frac{1}{1-\gamma
}w^{1-\gamma} \psi^{\gamma}(t)$. The key feature to note is that wealth $w$
and time $t$ are multiplicatively separable in $U$; thus, the optimal time to
annuitize one's wealth is independent of wealth and is, therefore, deterministic.

To solve the variational inequality for $\psi$, we hypothesize that the
``continuation region'' (i.e. the time when one does not annuitize) is of the
form $(0, T)$. In other words, one does not annuitize one's wealth until time
$T$. If our hypothesis is correct, then (\ref{8.3}) holds with equality on
$(0, T)$, while (\ref{8.4}) holds with inequality on $(0, T)$ and with
equality at $t = T$. Given any value of $T$, we can solve this boundary-value
problem; write $\phi= \phi(t; T)$ as the solution of this problem. For $t <
T$, the function $\phi$ is given by%

\begin{equation}
\phi(t;T)=\left(  \frac{\bar{a}_{x+T}^{S}}{\left(  \bar{a}_{x+T}^{O}\right)
^{1-\gamma}}\right)  ^{\frac{1}{\gamma}}e^{-\frac{r-\delta\left(
1-\gamma\right)  }{\gamma}\left(  T-t\right)  }\left(  _{T-t}p_{x+t}%
^{S}\right)  ^{\frac{1}{\gamma}}+\int_{t}^{T}e^{-\frac{r-\delta\left(
1-\gamma\right)  }{\gamma}\left(  s-t\right)  }\left(  _{s-t}p_{x+t}%
^{S}\right)  ^{\frac{1}{\gamma}}ds, \label{3.4}%
\end{equation}

\noindent with $\delta=r+ \frac{1}{2 \gamma} \left(  \frac{\mu-r}{\sigma}
\right)  ^{2}$. For $t \ge T$, we have%

\begin{equation}
\phi(t; T) = \left(  \frac{\bar{a}_{x+T}^{S}}{\left(  \bar{a}_{x+T}%
^{O}\right)  ^{1-\gamma}} \right)  ^{\frac{1}{\gamma}}. \label{3.4a}%
\end{equation}

The function $\psi$ is the \textit{smooth} solution of the variational
inequality in (\ref{8.3}) and (\ref{8.4}), for which there will be a
\textit{unique} $T$. (That is, $\phi$ will have a continuous first derivative
at $t = T \ge0$.) This optimal value of $T \ge0$ is the one such that
$\frac{1}{1 - \gamma} \phi^{\gamma}(t; T)$ is maximized. Note that it is also
possible that the optimal time to annuitize is right now (i.e. $T = 0$).

Define the function $\tilde U$ by $\tilde U(w, t; T) = \frac{1}{1-\gamma
}w^{1-\gamma} \phi^{\gamma}(t; T)$. Then, the value function $U$ in
(\ref{3.1}) is given by $U(w, t) = \max_{T \ge0} \tilde U(w, t; T)$, in which
the optimal value of $T$ is independent of wealth $w$ because $w$ factors from
the expression for $\tilde U$. In other words, the optimal time to annuitize
is \textit{deterministic} at time $t$.

One can obtain the optimal consumption and investment policies from the
first-order necessary conditions in (\ref{8.1}); they are given in feedback
form by%

\begin{equation}
C_{t}^{*}=c^{*}\left(  W_{t}^{*},t\right)  = \frac{W_{t}^{*}} {\psi(t)} ,
\label{3.5}%
\end{equation}

\noindent and%

\begin{equation}
\Pi_{t}^{*}=\pi^{*}\left(  W_{t}^{*},t\right)  =\frac{\mu-r}{\sigma^{2}\gamma}
\, W_{t}^{*} , \label{3.6}%
\end{equation}

\noindent respectively, in which $W_{t}^{\ast}$ is the optimally controlled
wealth before annuitization (time $T$). If we are in the case of logarithmic
utility one can show\footnote{See the earlier working paper version of this
paper, Milevsky and Young (2002a) for a proof and exploration of this fact.}
that the optimal consumption rate is $C_{t}^{\ast}=\frac{W_{t}^{\ast}}{\bar
{a}_{x+t}^{S}}.$

It is interesting to note that if $r=0$, in which case the denominator of the
optimal consumption, $\psi$, collapses to a (subjective) life expectancy, then
the consumption rate is precisely the minimum rate mandated by the U.S.
Internal Revenue Service for annual consumption withdrawals from IRAs after
age 71. Specifically, the proportion required to be withdrawn from one's
annuity each year equals the start-of-year balance divided by the future
expectation of life. Because $r>0$ in reality, the minimum IRS-mandated
consumption rate is less than what is optimal for individuals with logarithmic
utility and with mortality equal to that in the IRS tables.

To find the optimal time of annuitization, differentiate $\tilde U = \tilde
U(w, t; T)$ with respect to $T$, while assuming $t < T$. One can show that%

\begin{equation}
\frac{\delta\tilde U}{\delta T}\propto\left[  \frac{\gamma}{1-\gamma}\left(
\frac{\bar{a}_{x+T}^{S}}{\bar{a}_{x+T}^{O}}\right)  ^{-\frac{1-\gamma}{\gamma
}}-\frac{1}{1-\gamma}+\frac{\bar{a}_{x+T}^{S}}{\bar{a}_{x+T}^{O}}\right]  +
\bar{a}_{x+T}^{S}\left[  \delta-\left(  r+\lambda_{x+T}^{O}\right)  \right]  .
\label{3.7}%
\end{equation}

\noindent Thus, if the expression on the right-hand side of (\ref{3.7}) is
negative for all $T \ge0$, then it is optimal to annuitize one's wealth
immediately, and we have $U(w, t) = \tilde U(w, t; 0)$. However, if there
exists a value $T^{\ast}>0$ such that the right-hand side of (\ref{3.7}) is
positive for all $0\leq\ T<T^{\ast}$ and is negative for all $T>T^{\ast}$,
then it is optimal to annuitize one's wealth at time $T^{\ast}$, and we have
$U(w, t) = \tilde U(w, t; T^{*})$. In all the examples we present below, one
of these two conditions holds. We repeat that the decision to annuitize is
independent of one's wealth, an artifact of CRRA utility.

It is straightforward to show that $\tilde U(w, t; T)$ having a continuous
derivative at $t = T > 0$ means that $T$ is a \textit{critical point} of
$\tilde U$; that is, the right-hand side of (\ref{3.7}) is zero at that value
of $T$. Moreover, if that critical point $T = T^{*}$ \textit{maximizes}
$\tilde U(w, t; T)$, then (\ref{8.4}) holds with strict inequality on $(0,
T^{*})$, as desired.

If the subjective and objective forces of mortality are equal, then we have%

\begin{equation}
\frac{\delta\tilde U}{\delta T}\propto\left[  \delta-\left(  r+\lambda
_{x+T}\right)  \right]  . \label{3.8}%
\end{equation}

\noindent In this case, if the hazard rate $\lambda_{x+t}$ is increasing with
respect to time $t$, then \textit{either} $\delta\leq\left(  r+\lambda
_{x}\right)  $, from which it follows that it is optimal to annuitize one's
wealth immediately, \textit{or} $\delta>\left(  r+\lambda_{x}\right)  $, from
which it follows that there exists a time $T$ in the future (possibly
infinity) at which it is optimal to annuitize one's wealth. The optimal age to
purchase a fixed life annuity is when the force of mortality $\lambda_{x}$ is
greater than a constant (reminiscent of Merton's constant) defined by:%

\begin{equation}
M:=\frac{1}{2\gamma}\left(  \frac{\mu-r}{\sigma}\right)  ^{2}.
\end{equation}

\noindent One can then think of the hazard rate as a form of excess return on
the annuity due to the embedded mortality credits and the fact that liquid
wealth reverts to the insurance company when the buyer of the annuity dies.
This annuity purchase condition leads to a number of appealing insights.
Namely, higher levels of risk aversion ($\gamma$) and higher levels of
investment volatility ($\sigma$) lead to lower annuitization ages, since the
constant $M$ decreases under larger $\gamma,\sigma$ and increases under higher
levels of $\mu$.

We observe that if the subjective force of mortality is \textit{different}
than the objective force of mortality, then the optimal time of annuitization
increases from the $T$ given by the zero of the right-hand side of
(\ref{3.7}). We can show mathematically that this is true if the subjective
force of mortality varies from the objective force to the extent that $\bar
{a}_{x}^{S}<2\bar{a}_{x}^{O}$ for all ages $x$ (see Appendix A), and we
conjecture that it is true in general. Note that this inequality is
automatically true for people who are less healthy because in this case
$\bar{a}_{x}^{S}<\bar{a}_{x}^{O}$ for all $x$. For an individual who is less
healthy than the average person, the annuity will be too expensive, and the
person will want to delay annuitizing her wealth.

On the other hand, for an individual who is healthier than the average person,
the annuity will be relatively cheap. However, such a healthy person will live
longer on average and will be interested in receiving a larger annuity benefit
by consuming less now and by waiting to buy the annuity later in life.
Therefore, a healthy person is also willing to delay annuitizing her wealth in
exchange for a larger annuity benefit (for a longer time). Note that this
result is likely driven by the fact that in this model, we force the investor
to consume the totality of the income from the annuity. If the investor were
allowed to save some of that annuity income and invest it in the stock market,
this result would not necessarily hold.

Of course, by following the optimal policies of investment, consumption, and
annuitizing one's wealth, an individual runs the risk of being able to consume
less after annuitizing wealth than if she had annuitized wealth immediately at
time $t=0$. Naturally, there is the chance of the exact opposite, namely that
the lifetime annuity stream will be higher. Therefore, to quantify this risk,
we calculate the probability associated with various consumption outcomes. See
Appendix B for the formula of this probability. We include calculations of it
in a numerical example below.

Finally, we define a metric for measuring the loss in value from annuitizing
prematurely by computing the additional wealth that would be required to
compensate the utility maximizer for forced annuitization. This is akin to the
annuity equivalent wealth used by MPWB (1999), which we prefer to label a
subjective option value. Technically, it is defined to be the least amount of
money $h$ that when added to current wealth $w$ makes the person indifferent
between annuitizing now (with the extra wealth) and annuitizing at time $T$
(without the extra wealth). Thus, $h$ is given by%

\begin{equation}
U(w, t; T) = U(w+h,t;0), \label{3.9}%
\end{equation}

\noindent in which $T$ is the optimal time of annuitization. In the examples
in the next section, we express $h$ as a percentage of wealth $w$. This is
appropriate because $U$ exhibits CRRA with respect to $w$.

\section{Numerical Examples: Annuitize All or Nothing}

In this section, we present two numerical examples to illustrate the results
from the previous section. To start, although most mortality tables are
discretized, we require a continuous-time mortality law. We use a Gompertz
force of mortality, which is common in the actuarial literature for annuity
pricing. See Frees, Carriere, and Valdez (1996) for examples of this model in
annuity pricing. This model for mortality has also been employed in the
economics literature for pricing insurance; see Johansson (1996), for example.
The force of mortality is written $\lambda_{x}=\exp\left(  \left(  x-m\right)
/b\right)  /b$ in which $m$ is a modal value and $b$ is a scale parameter.
Note how the force of mortality itself increases exponentially with age.

In this paper, we fit the parameters of the Gompertz, namely $m$ and $b$, to
the Individual Annuity Mortality 2000 (basic) Table with projection scale G.
For males, we fit parameters $\left(  m,b\right)  =(88.18,10.5)$; for females,
($92.63,8.78).$ Initially, we assume that the subjective and objective forces
of mortality are equal. Throughout this section, we assume that the seller of
the annuity uses the female hazard rate to price annuities for women;
similarly, for men. Figure 1 shows the graph of the probability density
function of the future-lifetime random variable under a Gompertz hazard rate
that is fitted to the discrete mortality table.\footnote{We actually fit a
Makeham hazard rate, or force of mortality, namely $\lambda+\exp\left(
\left(  x-m\right)  /b\right)  /b$ in which $\lambda\geq0$ is a constant that
models an accident rate. However, the fitted value of $\lambda$ was $0$, so
the effective form of the hazard rate is Gompertz (Bowers et al., 1997).}%

\[%
\begin{tabular}
[c]{||c||}\hline\hline
\textbf{Figure 1 about here.}\\\hline\hline
\end{tabular}
\]

As for the capital market parameters, in both our examples, the risky stock is
assumed to have drift $\mu=0.12$ and volatility $\sigma$ $=0.20$. This is
roughly in line with numbers provided by Ibbotson Associates (2001), which are
widely used by practitioners when simulating long-term investment returns. We
assume that the nominal rate of return of the riskless bond is $r=0.06$. We
display values for the option to delay annuitization $h$, for three different
levels of risk aversion, $\gamma=1$ (logarithmic utility) and $\gamma=2$ and
$\gamma=5$. A variety of studies have estimated the value of $\gamma$ to lie
between 1 and 2. See, the paper by Friend and Blume (1975) that provides an
empirical justification for constant relative risk aversion, as well as the
more recent MPWB (1999) paper in which the CRRA value is taken between 1 and
2. In the context of estimating the present value of a variable annuity for
Social Security, Feldstein and Ranguelova (2001) provide some qualitative
arguments that the value of CRRA is less than 3 and probably even less than 2.
On the other hand, some of the equity premium literature, see Campbell and
Viceira (2002) suggests that risk aversion levels might be much higher, which
is why we have also displayed results for $\gamma=5$.

\subsection{Example \#1}

Table 1 provides the optimal age of annuitization -- and what we have labeled
the value of the option to delay as a percentage of initial wealth -- as well
as the probability of consuming less at the optimal time of annuitization than
if one had annuitized one's wealth immediately. We refer to this as the
probability of a deferral failure. \ We provide numerical results for both
males and females under very low ($\gamma=1$), low ($\gamma=2$) and high
($\gamma=5$) coefficients of relative risk aversion.%

\[%
\begin{tabular}
[c]{||c||}\hline\hline
\textbf{Table 1a about here.}\\\hline\hline
\end{tabular}
\]

Note that females annuitize at older ages compared to males because the
mortality rate of females is lower at each given age. Also, note that more
risk averse individuals wish to annuitize sooner, an intuitively pleasing
result. However, notice that even at relatively high ($\gamma=5$) levels of
risk aversion, males do not annuitize prior to age 63 and females do not
annuitize prior to age 70. Finally, the value of the option to delay
annuitization -- which is effectively the certainty equivalent of the welfare
loss from annuitizing immediately -- decreases as one gets closer to the
optimal age of annuitization, as one expects.%

\[%
\begin{tabular}
[c]{||c||}\hline\hline
\textbf{Table 1b about here.}\\\hline\hline
\end{tabular}
\]

The probability of deferral failure reported in Table 1a, although seemingly
high, is balanced by the probability of ending up with more than, say, 20\% of
the original annuity amount. For example, for a 70-year-old female with
$\gamma=2$, the probability of consuming at least 20\% more at the optimal age
of annuitization than if she were to annuitize immediately is 0.474.
Obviously, on a utility-adjusted basis this is a worthwhile trade-off as
evidenced by the behavior of the value function. See Table 2 for tabulations
of the probability that the individual consumes at least 20\% more at the
optimal age of annuitization than if he or she were to annuitize immediately,
for various ages and for $\gamma=1$ and $2$.%

\[%
\begin{tabular}
[c]{||c||}\hline\hline
\textbf{Table 2 about here.}\\\hline\hline
\end{tabular}
\]

These ``upside'' probabilities decrease as the optimal age of annuitization
approaches. Also, for a given age, they decrease as the CRRA increases. This
makes sense because a less risk-averse person is less willing to face a
distribution with a higher variance.

\subsection{Example \#2}

We continue the assumptions in the previous example as to the financial
market. We have a male aged 60 with $\gamma=2$, whose objective mortality
follows that from the previous example; that is, annuity prices are determined
based on the hazard rate given there. For this example, suppose that the
subjective force of mortality is a multiple of the objective force of
mortality; specifically, $\lambda_{x}^{S}=\left(  1+f\right)  \lambda_{x}^{O}%
$, in which $f$ ranges from $-1$ (immortal) to infinity (at death's door).
This transformation is called the \textit{proportional hazard} transformation
in actuarial science introduced by Wang (1996), and it is similar to the
transformation examined by Johansson (1996) in the economic context of the
economic value of increasing one's life expectancy.

In Table 3, we present the imputed value of the option to delay annuitization,
the optimal age of annuitization, the optimal rate of consumption before
annuitization (as a percentage of current wealth), and the rate of consumption
after annuitization (also, as a percentage of current wealth). For comparison,
if the male were to annuitize his wealth at age 60, the rate of consumption
would be 8.34\%. Also, the optimal proportion invested in the risky stock
before annuitization is 75\%.%

\[%
\begin{tabular}
[c]{||c||}\hline\hline
\textbf{Table 3 about here.}\\\hline\hline
\end{tabular}
\]

Note that as the 60-year-old male's subjective mortality gets closer to the
objective (pricing) mortality, then the optimal age of annuitization
decreases. It seems that the optimal age of annuitization will be a minimum
when the subjective and objective forces of mortality equal, at least for
increasing forces of mortality. We conjecture that this result is true in
general, but we only have a proof of it when $\bar{a}_{x}^{S}<2\bar{a}_{x}%
^{O}$; see Appendix A. Also, note that the consumption rate before
annuitization increases as the person becomes less healthy, as expected.

Compare these rates of consumption with 8.34\%, the rate of consumption if the
male were to annuitize his wealth immediately. We see that if the male is
healthy relative to the pricing force of mortality, then he is willing to
forego current consumption in exchange for greater consumption when he
annuitizes, at least up to $f=-0.4.$ Past that point, the optimal rate of
consumption before annuitization is greater than 8.34\%. For a 60-year-old
male with $f=-0.2$ (20\% more healthy than average), see Figure 2 for a graph
of the expected consumption rate as a percentage of \textit{initial} wealth.
We also graph the 25$^{th}$ and 75$^{th}$ percentiles of his random
consumption. This individual expects to live to age 84.4. Note that the
annuitant has roughly a 70\% chance of consuming more throughout the remaining
life compared to annuitizing at age 60.%

\[%
\begin{tabular}
[c]{||c||}\hline\hline
\textbf{Figure 2 about here.}\\\hline\hline
\end{tabular}
\]

\section{Unrestricted Market: Annuitize Anything Anytime}

In this section, we consider the optimal annuity-purchasing problem for an
individual who seeks to maximize her expected utility of lifetime consumption
and bequest. In Section 6.1, we allow the individual to have rather general
preferences, while in Section 6.2, we specialize to the case for which
preferences exhibit constant relative risk aversion. We allow the individual
to buy annuities in lump sums or continuously, whichever is optimal. Our
results are similar to those of Dixit and Pindyck (1994, pp 359ff). They
consider the problem of a firm's irreversible capacity expansion. For our
individual, annuity purchases are also irreversible, and this leads to the
similarity in results. Specifically, a discrete jump in wealth can only occur
at the initial instant (in our case, with a lump-sum purchase of an annuity;
in their case, with an initial investment of capital); thereafter, the annuity
income remains constant or increases incrementally to keep wealth below a
given barrier (for Dixit and Pindyck, capital stock was either constant or
changed incrementally). In other words, the optimal control is a ``barrier
control'' policy.

In Section 6.2.1, we continue with CRRA preferences and linearize the HJB
equation in the region of no-annuity purchasing via a convex dual
transformation in the case for which there is no bequest motive. In Section
6.2.1.1, we provide an implicit analytical solution to the optimal annuity
purchasing problem developed in Section 6.2.1 in the case for which the force
of mortality is constant. This leads us to Section 7, which provides a full
set of numerical results.

\subsection{General Utility of Consumption and Bequest}

In this section, we show that the individual's optimal annuity purchasing is
given by a barrier policy in that she will annuitize just enough of her wealth
to stay on one side of the barrier in wealth-annuity space. In equation
(\ref{2.2}), we described the dynamics of the wealth for this individual.
Denote the random time of death of our individual by $\tau_{d}$. We assume
that $\tau_{d}$ is independent of the randomness in the financial market,
namely the Brownian motion $B$ driving the stock price. Thus, her value
function at time $t$, or at age $x+t$, is given by%

\begin{align}
&  U(w,A,t)\nonumber\\
&  =\sup_{\{c_{s},\pi_{s},A_{s}\}}\mathbf{E}^{w,A,t}\left[  \int_{t}^{\infty
}e^{-r(s-t)}\left.  _{s-t}p_{x+t}^{S}\right.  u_{1}(c_{s})ds+e^{-r(\tau
_{d}-t)}\,u_{2}(W_{\tau_{d}})\right] \nonumber\\
&  =\sup_{\{c_{s},\pi_{s},A_{s}\}}\mathbf{E}^{w,A,t}\left[  \int_{t}^{\infty
}e^{-r\left(  s-t\right)  }\left.  _{s-t}p_{x+t} ^{S}\right.  \left\{
u_{1}\left(  c_{s}\right)  +\lambda_{x+s}^{S} u_{2}\left(  W_{s-}\right)
\right\}  ds\right]  , \label{4.1}%
\end{align}

\noindent in which $u_{1}$ and $u_{2}$ are strictly increasing, concave
utility functions of consumption and bequest, respectively. Also,
$\mathbf{E}^{w,A,t}$ denotes the expectation conditional on $W_{t-}=w$ and
$A_{t-}=A$. In the last equality, we used the independence of $\tau_{d}$ from
the Brownian motion $B$ to simplify the expression for $U$. Note that we
assume the individual discounts future consumption at the riskless rate $r$
since the mortality discounting -- which increases the effective discount rate
-- is incorporated separately. The value function $U$ is jointly
concave\footnote{Due to space constraints, we refer the interested reader to
the earlier working paper version by Milevsky and Young (2002b) for a detailed
discussion of this and other properties of $U$.} in $w$ and $A$. 

We continue with a formal discussion of the derivation of the associated HJB
equation. Suppose that at the point $(w, A, t)$, it is optimal \textit{not} to
purchase any annuities. It follows from It\^{o}'s lemma that $U$ satisfies the
equation at $(w, A, t)$ given by%

\begin{align}
&  (r+\lambda_{x+t}^{S})U\nonumber\\
&  =U_{t}+(rw+A)U_{w}+\max_{\pi}\left[  \frac{1}{2}\sigma^{2}\pi^{2}
U_{ww}+\left(  \mu-r\right)  \pi U_{w}\right]  +\max_{c\geq0}\left[
-cU_{w}+u_{1}\left(  c\right)  \right]  +\lambda_{x+t}^{S}u_{2}\left(
w\right)  . \label{4.2}%
\end{align}

\noindent Because the above policy is in general suboptimal, \eqref{4.2} holds
as an inequality; that is, for all $(w, A, t)$,%

\begin{align}
&  (r+\lambda_{x+t}^{S})U\nonumber\\
&  \ge U_{t}+(rw+A)U_{w}+\max_{\pi}\left[  \frac{1}{2}\sigma^{2}\pi^{2}
U_{ww}+\left(  \mu-r\right)  \pi U_{w}\right]  +\max_{c\geq0}\left[
-cU_{w}+u_{1}\left(  c\right)  \right]  +\lambda_{x+t}^{S}u_{2}\left(
w\right)  . \label{4.3}%
\end{align}

Next, assume that at the point $(w,A,t)$ it \textit{is} optimal to buy an
annuity instantaneously. In other words, assume that the investor moves
instantly from $(w,A,t)$ to $(w-\bar{a}_{x+t}^{O}\Delta A,A+\Delta A,t)$.
Then, the optimality of this decision implies that%

\begin{equation}
U(w, A, t) = U(w - \bar{a}_{x+t}^{O} \Delta A, A + \Delta A, t), \label{4.4}%
\end{equation}

\noindent which in turns yields%

\begin{equation}
U_{A}(w, A, t) - \bar{a}_{x+t}^{O} U_{w}(w, A, t) = 0. \label{4.5}%
\end{equation}

\noindent Note that the lump-sum purchase is such that the marginal utility of
annuity income equals the adjusted marginal utility of wealth, in which we
adjust the marginal utility of wealth by multiplying by the cost of $\$1$ of
annuity income. This result parallels many such in economics. Indeed, the
marginal utility of annuity income is the marginal utility of the benefit,
while the adjusted marginal utility of wealth is the marginal utility of the
cost. Thus, the lump-sum purchase is such that the marginal utilities are equated.

However, such a policy is in general suboptimal; therefore, \eqref{4.4} holds
as an inequality and \eqref{4.5} becomes%

\begin{equation}
\bar{a}_{x+t}^{O}U_{w}(w,A,t)-U_{A}(w,A,t)\geq0. \label{4.6}%
\end{equation}

By combining \eqref{4.3} and \eqref{4.6}, we obtain the HJB equation
\eqref{4.7} below associated with the value function $U$ given in \eqref{4.1}.
The following result can be proved as in Zariphopoulou (1992), for example.

\medskip

\textbf{Proposition 6.1:} \textit{The value function U is a constrained
viscosity solution of the Hamilton-Jacobi-Belman equation}%

\begin{equation}%
\begin{array}
[c]{c}%
\min\left[  (r+\lambda_{x+t}^{S}) U-U_{t} -(rw+A)U_{w}-\max_{\pi}\left(
\frac{1}{2}\sigma^{2}\pi^{2}U_{ww}+ (\mu-r) \pi U_{w}\right)  \right. \\
\left.  -\max_{c\geq0}\left(  -cU_{w}+u_{1} \left(  c \right)  \right)
-\lambda_{x+t}^{S} u_{2} \left(  w \right)  , \hbox{ } \bar a_{x+t}^{O}
U_{w}-U_{A} \right]  = 0.
\end{array}
\label{4.7}%
\end{equation}

\medskip

Equation \eqref{4.5} defines a ``barrier" in wealth-annuity income space. If
wealth and annuity income lie to the right of the barrier at time $t$, then
the individual will immediately spend a lump sum of wealth to move diagonally
to the barrier (up and to the left). The move is diagonal because as wealth
decreases to purchase more annuities, annuity income increases. Thereafter,
annuity income is either constant if wealth is low enough to keep to the left
of the barrier, or annuity income responds continuously to infinitesimally
small changes of wealth at the barrier.

Thus, as in Dixit and Pindyck (1994, pp 359ff) or in Zariphopoulou (1992), we
have discovered that the optimal annuity-purchasing scheme is a type of
barrier control. Other barrier control policies appear in finance and
insurance. In finance, Zariphopoulou (1999, 2001) reviews the role of barrier
policies in optimal investment in the presence of transaction costs; also see
the references within her two articles. See Gerber (1979) for a classic text
on risk theory in which he includes a section on optimal dividend payout and
shows that it follows a type of barrier control.

\subsection{Constant Relative Risk Aversion Preferences}

In this subsection, we specialize the results of the previous subsection to
the case for which the individual's preferences exhibit CRRA. For this case,
we can reduce the problem by one dimension, and we show that the barrier given
in the previous section is a ray emanating from the origin in wealth-annuity
space. Let%

\begin{equation}
u_{1}(c)=\frac{c^{1-\gamma}}{1 - \gamma}, \text{ and } u_{2}(w) = k u_{1}(w),
\quad\gamma>0, \gamma\ne1, k \ge0. \label{4.9}%
\end{equation}

\noindent The parameter $k \ge0$ weights the utility of bequest relative to
the utility of consumption. Davis and Norman (1990) and Shreve and Soner
(1994) show that for CRRA preferences in the problem of consumption and
investment in the presence of transaction costs, the value function $U$ is a
solution of its HJB equation in the classical sense, not just in the viscosity
sense. Generally, if the force of mortality is ``eventually" large enough to
make the value function well-defined, then this result holds for our problem, too.

For the utility functions in ({\ref{4.9}), it turns out that the value
function $U$ is homogeneous of degree $1-\gamma$ with respect to wealth $w$
and annuity income $A$. That is, $U(bw, bA, t) = b^{1-\gamma}U(w,A,t)$ for
$b>0$. Thus, if we define $V$ by $V(z,t) = U(z,1,t)$, then we can recover $U$
from $V$ by }%

\begin{equation}
U(w,A,t)=A^{1-\gamma}V(w/A,t), \text{ for }A>0.
\end{equation}

\noindent It follows that the HJB equation for $U$ from Proposition 6.1
becomes the following equation for $V$:%

\begin{align}
&  \min\left[  {}\right.  (r+\lambda_{x+t}^{S})V-V_{t}-(rz+1)V_{z}-\max
_{\hat{\pi}}\left(  \frac{1}{2}\sigma^{2}\hat{\pi}^{2}V_{zz}+(\mu-r)\hat{\pi
}V_{z}\right)  \nonumber\\
&  \qquad-\max_{\hat{c}\geq0}\left(  -\hat{c}V_{z}+\frac{\hat{c}^{1-\gamma}%
}{1-\gamma}\right)  -k\lambda_{x+t}^{S}\frac{z^{1-\gamma}}{1-\gamma}%
,\quad(z+\overline{a}_{x+t}^{O})V_{z}-(1-\gamma)V\left.  {}\right]
=0,\label{4.10}%
\end{align}

\noindent in which $\hat{c}=c/A$, and $\hat{\pi}=\pi/A$. Davis and Norman
(1990) and Shreve and Soner (1994) use the same transformation in the problem
of consumption and investment in the presence of transaction costs. Also,
Duffie and Zariphopoulou (1993) and Koo (1998) use this transformation to
study optimal consumption and investment with stochastic income.

Due to space considerations we simply refer to the working paper version by
Milevsky and Young (2002b) which study properties of the optimal consumption
and investment policies. Please refer to that work for details on the proof of
the following proposition that describes the actions of the individual.

\textbf{Proposition 6.2:} \textit{For each value of }$t\geq0$\textit{, there
exists a value of the wealth-to-income ratio $z_{0}(t)$ that solves}
\begin{equation}
(z_{0}(t) + \bar{a}_{x+t}^{O}) V_{z}(z_{0}(t), t) = (1 - \gamma) V(z_{0}(t),
t),
\end{equation}
\noindent\textit{such that}

(i) \quad\textit{If }$z = w/A >z_{0}(t)$\textit{, then the individual
immediately buys an annuity so that}
\begin{equation}
\frac{w-\Delta A \bar a_{x+t}^{O}}{A+\Delta A} = z_{0}(t);
\end{equation}
\noindent\textit{Thus, $V(z, t) = V(z_{0}(t), t)$ in this case.}

(ii) \quad\textit{If }$z = w/A <z_{0}(t)$\textit{, then the individual buys no
annuity; i.e., she is in the region of inaction. Thus, in this case, $V$
solves}
\begin{align}
&  (r+\lambda_{x+t}^{S})V\label{4.8}\\
&  =V_{t}+(rz+1)V_{z}+\max_{\hat{\pi}}\left(  \frac{1}{2}\sigma^{2}\hat{\pi
}^{2}V_{zz}+(\mu- r) \hat{\pi}V_{z}\right)  +\max_{\hat{c}\geq0}\left(
-\hat{c}V_{z}+\frac{\hat{c}^{1-\gamma} }{1-\gamma}\right)  +k\lambda_{x+t}%
^{S}\frac{z^{1-\gamma}}{1-\gamma}.\nonumber
\end{align}
\noindent\textit{It follows that at each time point, the barrier $w =
z_{0}(t)A$ is a ray emanating from the origin and lying in the first quadrant
of }$(w,A)$\textit{ space.}

\medskip

Note that if $z_{0}(t) < \infty$, then it is optimal for the individual to
have \textit{positive} annuity income because the positive $w$ axis lies in
the region $\{(w, A, t): w/A < z_{0}(t)\}$.

Davis and Norman (1990) and Shreve and Soner (1994) find results similar to
those in Proposition 6.2 for the problem of optimal consumption and investment
in the presence of proportional transaction costs. In the next subsection, we
show how to linearize the HJB equation of the individual who has no bequest motive.

\subsubsection{Zero Bequest Motive: Linearization of the HJB Equation}

Up until now we have assumed both utility of bequest and consumption in our
specification. In this subsection, we linearize the nonlinear partial
differential equation for $V$ in the region of inaction given by equation
(\ref{4.8}) with no bequest motive ($k=0$). To this end, we consider the
convex dual of $V$ defined by%

\begin{equation}
\tilde{V}(y,t)= \max_{z>0}\left[  V(z,t)-zy \right]  . \label{4.11}%
\end{equation}

\noindent The critical value $z^{*}$ solves the equation $0=V_{z}(z,t)-y$;
thus, $z^{*}=I(y,t)$, in which $I$ is the inverse of $V_{z}$ with respect to
$z$. Note that one can retrieve the function $V$ from $\tilde{V}$ by the relationship%

\begin{equation}
V(z,t)= \min_{y>0} \left[  \tilde{V}(y,t)+zy \right]  . \label{4.14}%
\end{equation}

\noindent Indeed, the critical value $y^{*}$ solves the equation $0=\tilde
{V}_{y}(y,t)+z=-I(y,t)+z$; thus, $y^{*}=V_{z}(z,t)$.

In the partial differential equation for $V$ with no bequest motive ($k=0$),
let $z=I(y,t)$ and rewrite the equation in terms of $\tilde{V}$ to obtain%

\begin{equation}
\tilde{V}_{t}-(r+\lambda_{x+t}^{S})\tilde{V}+\lambda_{x+t}^{S}y\tilde{V}%
_{y}+my^{2}\tilde{V}_{yy}=-y-\frac{\gamma}{1-\gamma}y^{1-\frac{1}{\gamma}},
\label{4.17}%
\end{equation}

\noindent in which $m=\frac{1}{2} \left(  \frac{\mu-r}{\sigma} \right)  ^{2}.$
Note that (\ref{4.17}) is a \textit{linear} partial differential equation.

Next, consider the boundary condition $U_{A}(w,A,t)=\bar a_{x+t}^{O} \,
U_{w}(w,A,t)$ from equation (\ref{4.5}). In terms of $V$, this condition can
be written as in equation (27), and we repeat it here for convenience%

\begin{equation}
-(1-\gamma)V(z_{0}(t),t)+(z_{0}(t)+\bar a_{x+t}^{O})V_{z}(z_{0}(t),t)=0.
\label{4.18}%
\end{equation}

\noindent Smooth pasting at the boundary implies that the derivative of this
boundary condition with respect to $z$ evaluated at $z=z_{0}(t)$ holds and is
given by%

\begin{equation}
\gamma V_{z}(z_{0}(t),t)+(z_{0}(t)+\bar a_{x+t}^{O})V_{zz}(z_{0}(t),t)=0.
\label{4.19}%
\end{equation}

We also have a boundary condition at $z=0$ because at that point, the
individual has no wealth to invest in the risky asset. Write $\hat\pi^{*}$ in
terms of $\tilde V$: $\hat\pi^{*}(y, t) = {\frac{\mu- r }{\sigma^{2}}} y
\tilde V_{yy}$. Thus, for $z = 0$ (with the corresponding value for $y$
written $y_{a}(t)$), we have that either $y_{a}(t) = 0$ or $\tilde
V_{yy}(y_{a}(t), t) = 0$.

Because $V_{z}>0$ is strictly decreasing with respect to $z$, we have
$y_{a}(t) > y_{0}(t) \ge0$ for all $t \ge0$, in which $y_{a}(t)$ and
$y_{0}(t)$ are defined by%

\begin{equation}
y_{a}(t)=V_{z}(0,t), \hbox{ and } y_{0}(t)=V_{z}(z_{0}(t),t). \label{4.20}%
\end{equation}

\noindent Thus, because $y_{a}(t) > 0$, in terms of $\tilde{V}$, the boundary
conditions become%

\begin{equation}
\tilde{V}_{y}(y_{a}(t),t)=0, \label{4.22}%
\end{equation}

\noindent for%

\begin{equation}
\tilde{V}_{yy}(y_{a}(t),t)=0, \label{4.23}%
\end{equation}

\noindent and%

\begin{equation}
(1-\gamma)\tilde{V}(y_{0}(t),t)+\gamma y_{0}(t) \tilde{V}_{y}(y_{0}(t),t)=\bar
a_{x+t}^{O} \, y_{0}(t), \label{4.24}%
\end{equation}

\noindent for%

\begin{equation}
\tilde{V}_{y}(y_{0}(t),t)+\gamma y_{0}(t) \tilde{V}_{yy}(y_{0}(t),t)=\bar
a_{x+t}^{O}. \label{4.25}%
\end{equation}

\medskip

\noindent\textbf{6.2.1.1 \quad Constant Force of Mortality}

While still operating within the zero bequest world, if we assume that the
forces of mortality are constant, that is, $\lambda_{x+t}^{S}\equiv\lambda
^{S}$ and $\lambda_{x+t}^{O}\equiv\lambda^{O}$ for all $t\geq0$, then we can
obtain an \textquotedblleft implicit\textquotedblright\ analytical solution of
the value function $V$ via the boundary-value problem given by (\ref{4.17})
and (\ref{4.22}) - (\ref{4.25}). See Neuberger (2003) for recent and related
work. In this case, $V$, $\tilde{V}$, $y_{a}$, and $y_{0}$ are independent of
time, so (\ref{4.17}) becomes the ordinary differential equation%

\begin{equation}
-(r+\lambda^{S})\tilde{V}\left(  y\right)  +\lambda^{S}y\tilde{V}^{\prime
}\left(  y\right)  + y^{2}\tilde{V}^{\prime\prime} \left(  y\right)
=-y-\frac{\gamma}{1-\gamma}y^{1-\frac{1}{\gamma}}, \label{4.26}%
\end{equation}

\noindent with boundary conditions%

\begin{equation}
\tilde{V}^{\prime\prime}(y_{a})=0, \label{4.27}%
\end{equation}

\noindent for%

\begin{equation}
\tilde{V}^{\prime}(y_{a})=0, \label{4.28}%
\end{equation}

\noindent and%

\begin{equation}
(1-\gamma)\tilde{V}(y_{0})+\gamma y_{0} \tilde{V}^{\prime}(y_{0})=\frac{y_{0}%
}{r+\lambda^{O}}, \label{4.29}%
\end{equation}

\noindent for%

\begin{equation}
\tilde{V}^{\prime}(y_{0})+\gamma y_{0} \tilde{V}^{\prime\prime}(y_{0}%
)=\frac{1}{r+\lambda^{O}}. \label{4.30}%
\end{equation}

The general solution of (\ref{4.26}) is%

\begin{equation}
\tilde{V}(y)=D_{1} y^{B_{1}} + D_{2} y^{B_{2}} + \frac{y}{r} + C_{2} y^{1 -
\frac{1}{\gamma}}, \label{4.31}%
\end{equation}

\noindent with $D_{1}$ and $D_{2}$ constants to be determined by the boundary
conditions, with $C_{2}$ given by%

\begin{equation}
C_{2}=r + \frac{\lambda^{S}}{\gamma} - m \frac{1-\gamma}{\gamma^{2}},
\label{4.32}%
\end{equation}

\noindent with $B_{1}$ and $B_{2}$ given by%

\begin{equation}
B_{1} = \frac{1}{2m} \left[  (m - \lambda^{S}) + \sqrt{(m - \lambda^{S})^{2} +
4m(r + \lambda^{S})} \right]  > 1, \label{4.33}%
\end{equation}

\noindent and%

\begin{equation}
B_{2} = \frac{1}{2m} \left[  (m - \lambda^{S}) - \sqrt{(m - \lambda^{S})^{2} +
4m(r + \lambda^{S})} \right]  < 0. \label{4.34}%
\end{equation}

The boundary conditions at $y_{0}$ give us%

\begin{equation}
D_{1} \{1+ \gamma(B_{1} - 1) \} y_{0}^{B_{1}} + D_{2} \{1+ \gamma(B_{2} - 1)\}
y_{0}^{B_{2}} + \frac{y_{0}}{r} = \frac{y_{0}}{r+\lambda^{O}}, \label{4.35}%
\end{equation}

\noindent and%

\begin{equation}
D_{1} B_{1} \{1+ \gamma(B_{1} - 1) \} y_{0}^{B_{1}} + D_{2} B_{2} \{1+
\gamma(B_{2} - 1) \} y_{0}^{B_{2}} + \frac{y_{0}}{r} = \frac{y_{0}}
{r+\lambda^{O}}. \label{4.36}%
\end{equation}

\noindent Solve equations (\ref{4.35}) and (\ref{4.36}) to get $D_{1}$ and
$D_{2}$ in terms of $y_{0}$:%

\begin{equation}
D_{1} = - \frac{\lambda^{O}}{r(r + \lambda^{O})} \frac{1 - B_{2}}{B_{1} -
B_{2}} \frac{y_{0}^{1 - B_{1}}}{1+ \gamma(B_{1} - 1)}, \label{4.37}%
\end{equation}

\noindent and%

\begin{equation}
D_{2} = - \frac{\lambda^{O}}{r(r + \lambda^{O})} \frac{B_{1} - 1}{B_{1} -
B_{2}} \frac{y_{0}^{1 - B_{2}}}{1+ \gamma(B_{2} - 1)}. \label{4.38}%
\end{equation}

Next, substitute for $D_{1}$ and $D_{2}$ in $\tilde V^{\prime}(y_{a}) + \gamma
y_{a} \tilde V^{\prime\prime}(y_{a}) = 0$ from (\ref{4.27}) and (\ref{4.28})
to get%

\begin{equation}
\frac{\lambda^{O}}{r+\lambda^{O}} \frac{B_{1} (1 - B_{2})}{B_{1} - B_{2}}
\left(  \frac{y_{a}}{y_{0}} \right)  ^{B_{1} - 1} + \frac{\lambda^{O}
}{r+\lambda^{O}} \frac{B_{2} (B_{1} - 1)}{B_{1} - B_{2}} \left(  \frac{y_{a}%
}{y_{0}} \right)  ^{B_{2} - 1} = 1. \label{4.39}%
\end{equation}

\noindent(\ref{4.39}) gives us an equation for the ratio $y_{a}/y_{0} > 1$. To
check that (\ref{4.39}) has a unique solution greater than 1, note that the
left-hand side (i) equals $\lambda^{O}/(r+\lambda^{O}) < 1$ when we set
$y_{a}/y_{0} = 1$, (ii) goes to infinity as $y_{a}/y_{0}$ goes to infinity,
and (iii) is strictly increasing with respect to $y_{a}/y_{0}$.

Next, substitute for $D_{1}$ and $D_{2}$ in $\tilde V^{\prime}(y_{a}) = 0$
from (\ref{4.27}) to get%

\begin{align}
- \frac{\lambda^{O}}{r(r+\lambda^{O})} \frac{B_{1}(1 - B_{2})}{B_{1} - B_{2}}
&  \frac{(y_{a}/y_{0})^{B_{1}-1}}{1 + \gamma(B_{1} - 1)} - \frac{\lambda^{O}%
}{r(r+\lambda^{O})} \frac{B_{2}(B_{1} - 1)}{B_{1} - B_{2}} \frac{(y_{a}%
/y_{0})^{B_{2}-1}}{1 + \gamma(B_{2} - 1)}\nonumber\\
&  + \frac{1}{r} + C_{2} \left(  1 - \frac{1}{\gamma} \right)  y_{a}%
^{-\frac{1}{\gamma}} = 0. \label{4.40}%
\end{align}

\noindent Substitute for $y_{a}/y_{0}$ in equation (\ref{4.40}), and solve for
$y_{a}$. Finally, we can get $y_{0}$ from%

\begin{equation}
y_{0} = \frac{y_{a}}{y_{a}/y_{0}}, \label{4.41}%
\end{equation}

\noindent and $D_{1}$ and $D_{2}$ from equations (\ref{4.37}) and
(\ref{4.38}), respectively.

Once we have the solution for $\tilde{V}$, we can recover $V$ from%

\begin{align}
V(z)  &  =\max_{y>0}\left[  \tilde{V}(y)+zy\right] \nonumber\\
&  =\max_{y>0}\left[  D_{1}y^{B_{1}}+D_{2}y^{B_{2}}+\frac{y}{r}+C_{2}%
y^{1-\frac{1}{\gamma}}+zy\right]  , \label{4.42}%
\end{align}

\noindent in which the critical value $y^{\ast}$ solves%

\begin{equation}
D_{1}B_{1}y^{B_{1}-1}+D_{2}B_{2}y^{B_{2}-1}+\frac{1}{r}+C_{2}\left(
1-\frac{1}{\gamma}\right)  y^{-\frac{1}{\gamma}}+z=0. \label{4.43}%
\end{equation}

\noindent Thus, for a given value of $z=w/A$, solve (\ref{4.43}) for $y$ and
substitute that value of $y$ into (\ref{4.42}) to get $U(w,A)=V(z)$. Perhaps
more importantly, we are interested in the critical value $z_{0}$ above which
an individual spends a lump sum to purchase more annuity income. We pursue
this in the examples in the next section.

\section{Numerical Examples: Annnuitize Anything Anytime}

\noindent In this section, we provide a variety of numerical examples to
illustrate the results of our \emph{anything anytime} model. We focus
attention on the impact of risk aversion, investment volatility, and insurance
fees on the optimal amount annuitized.

In the first set of results, we assumed the following values for the hazard
rate parameters: $\lambda^{S}=\lambda^{O}=0.04$. That is, the force of
mortality is constant and therefore the expected future lifetime is:
$1/\lambda=25$ years. Furthermore, we set the risk-free interest rate to be
$r=0.04$, the drift of the risky asset is $\mu=0.08$, and its volatility is
$\sigma=0.20$. We have selected these numbers -- which are lower than those
used in the earlier examples -- to better capture a real (after-inflation)
case in which Social Security benefits would be considered as part of the
pre-existing annuity.

In Table 4a, for various values of $\gamma$, we give the critical value of the
ratio of wealth to annuity income $z_{0}=w/A$ above which the individual will
spend a lump sum of wealth to increase her annuity income. We also include the
amount that the individual will spend on annuities for a given (pre-existing)
annuity income of $A=\$25,000$, namely $(w-z_{0}A)/(1+(r+\lambda^{O})z_{0})$.%

\[%
\begin{tabular}
[c]{||c||}\hline\hline
\textbf{Table 4a about here.}\\\hline\hline
\end{tabular}
\]

For example, a retiree with \$1,000,000 in liquid investable assets and with
\$25,000 in pre-existing annuity income would immediately (and irreversibly)
annuitize between \$727,620 and \$914,176 depending on the coefficient of
relative risk aversion. Notice from the same Table 4a that the amount spent on
annuities increases for a given level of wealth as the individual becomes more
risk averse, which is an intuitively pleasing result. Also, for a given level
of risk aversion, the amount spent on annuities decreases as wealth decreases.
In Table 4b, we present the results for the case when $A=\$50,000$. Notice
that when the pre-existing annuity income increases from \$40,000 to \$50,000,
but under the same level of wealth, the amount spent immediately on additional
annuity purchases is less.%

\[%
\begin{tabular}
[c]{||c||}\hline\hline
\textbf{Table 4b about here.}\\\hline\hline
\end{tabular}
\]

In fact, in this case someone with \$100,000 of wealth, or less, will not
annuitize any additional wealth when their coefficient of relative risk
aversion is $\gamma=2$ or less. We emphasize -- once again -- that these
numerical results assume that there is absolutely no bequest motive and no
importance placed in inheritance, a spouse or any other survivors. In the
presence of some weight on bequest (which would be expected in the real world)
the amount annuitized would not be any higher.%

\[%
\begin{tabular}
[c]{||c||}\hline\hline
\textbf{Table 4c about here.}\\\hline\hline
\end{tabular}
\]

Table 4c looks at a different aspect of the problem, namely the impact of
investment volatility $\sigma$ on the optimal amount annuitized in the
\emph{anything anytime} environment. In this case, we assume that same
$\lambda^{O}=\lambda^{S}=0.04$ hazard rate, $r=0.05$ risk free rate and
$\mu=0.12$ drift for the risky asset. We provide results under two differing
levels of risk aversion: high ($\gamma=5$) and low ($\gamma=2$). In both of
these cases, we assume a retiree/investor with \$1,000,000 of liquid
investable wealth and \$40,000 in pre-existing annuity income.

Notice from Table 4c that as the investment volatility $\sigma$ increases from
$0.12$ to $0.20$, the amount annuitized declines under both (and we can show,
all) levels of risk aversion. In fact, under high levels of volatility the
retiree/investor annuitizes \$472,871 for $\gamma=2$ and \$768,568 for
$\gamma=5$. These numbers drop to \$12,692 and \$469,789, respectively, when
the investment volatility is reduced from 0.20 to 0.12. The economic intuition
for this result is quite clear. As the relative risk of investing in the
high-return alternative declines, it becomes much less appealing, on a
risk-adjusted basis, to annuitize one's wealth.%

\[%
\begin{tabular}
[c]{||c||}\hline\hline
\textbf{Table 4d about here.}\\\hline\hline
\end{tabular}
\]

Table 4d investigates the impact of subjective vs. objective health status on
the amount annuitized in the same \emph{anything anytime} framework. In this
case we assume an objective (annuity pricing) hazard rate of $\lambda
^{O}=0.04$, but vary the subjective $\lambda^{S}$ from a value of 0.03 to
0.055. Thus, the insurance company believes the annuitant has a life
expectancy of $1/0.04=25.0$ years, while the annuitant believes they are
healthier (with a life expectancy increased to $1/0.03=33.3$ years) or
unhealthier (with a life expectancy decreased to $1/0.055=18.2$ years). One
can think of this as representing the impact of adverse selection or asymmetry
of information on the amount annuitized. As in Table 4c, we assume the
retiree/investor has \$1,000,000 in liquid wealth and pre-existing annuity
income of \$40,000 per annum. The market parameters are $r=0.05$ for the risk
free rate, $\mu=0.10$ drift of the risky asset and $\sigma=0.16$ for the
investment volatility.

In this case, and in contrast to the results from the \emph{all-or-nothing}
results, an increasing hazard rate has a uniform impact on the amount
annuitized. Namely, people who consider themselves to be in worse health
annuitize less, all else being equal. More specifically, a $\gamma=2$
retiree/investor who believes he has a life expectancy of (only) 18.2 years
will annuitize \$514,496 in contrast to \$574,840 when he believes his life
expectancy is a higher 33.3 years. The same result is observed at higher
levels of risk aversion, although it is less extreme in nominal (and marginal)
terms. Notice that a 15-year difference in perceived life expectancy reduces
the amount annuitized by \$60,000 under low ($\gamma=2$) levels of risk
aversion. But, at higher levels of ($\gamma=5$) risk aversion, the difference
is a mere \$28,000.

We remind the reader that these results have been obtained under a variety of
assumptions, namely a constant hazard rate (exponential future lifetime
distribution) and zero bequest motive, as well as constant risk-free rate and
risk-premia. To eliminate these restrictions is the subject of ongoing research.

\section{Conclusion and Main Insights}

Our paper locates the optimal dynamic policy of a utility-maximizing
individual who is interested in incorporating lifetime payout annuities (or
defined benefit pension income) into his or her retirement portfolio. We
investigated a variety of institutional arrangements and market structures
with differing restrictions and constraints. Our main results can be stated as follows.

\begin{itemize}
\item An individual who is faced with the irreversible decision of when to
start a fixed (nominal or real) lifetime pension annuity -- with the proviso
that annuitization must take place in an `all-or-nothing' format -- is endowed
with an incentive to delay that is quite valuable at younger ages.

\item There is an incentive to delay annuitization even in the absence of
bequest motives. This is because market imperfections do not allow retirees to
purchase payout annuities that offer complete asset allocation flexibility to
match ones subjective consumption preferences \emph{vis a vis} their health
status. Stated differently, a market in which instantaneously-renegotiated
life-contingent tontines are available would not give rise to our `option to
delay' result.

\item Under an all-or-nothing framework, which is a feature of many public and
private pension systems, the optimal age to annuitize is the age at which the
option to delay has zero time value. This value is defined equal to the loss
in utility that comes from not being able to behave optimally. This value
depends on a person's coefficient of relative risk aversion, as well as their
subjective health status.

\item Using historical market parameters and realistic mortality estimates, we
conclude that in this all-or-nothing framework the optimal age to purchase a
pure life-contingent annuity does not occur prior to age 70. This result is
consistent with a variety of probabilistic-based models which are based on the
relationship between mortality credits and the returns from competing asset
classes.\footnote{This after-age-70 result has also been advocated in a
variety of popular-press article, such as a recent piece in the \emph{Wall
Street Journal}, 3 September 2003, page C1, \textquotedblleft It Pays to
Delay: The Longer You Wait To Buy an Annuity, the More You
Get,\textquotedblright\ by Jonathan Clements.}

\item In the event that complete asset allocation flexibility is available
within the payout annuity, which is akin to variable immediate annuities that
are available in some countries and jurisdictions, the optimal age to
annuitize is indeed earlier. Of course, high management fees and expenses
relative to non-annuitized wealth can have a strong mitigating impact on the
benefits from annuitization.

\item When we move towards an open institutional system in which annuitization
can take place in small portions and at anytime, we find that
utility-maximizing investors should acquire a base amount of annuity income
(i.e. Social Security or a DB pension) and then annuitize additional amounts
if and when their wealth-to-income ratio exceeds a certain level. In this case
which we label \emph{anything anytime}, individuals annuitize a fraction of
wealth as soon as they have opportunity to do so -- i.e. they do \emph{not}
wait -- and they then purchase more annuities as they become wealthier.

\item Thus, in contrast to the all-or-nothing pension structure, in the case
of an open system where annuities can be purchased on an ongoing basis we find
that individuals prior to age 70 \textit{should} have a minimal amount of
annuity income and should immediately annuitize a fraction of wealth to create
this base level of lifetime income if they do not already have this from
pre-existing DB pensions. We reiterate that individuals should \textit{always}
hold some annuities, even in the presence of a bequest motive, as long as
$z_{0}(t)$ in Proposition 6.2 is less than infinity.

\item Finally, our \emph{anything anytime} model indicates that ceteris
paribus, a larger amount of wealth relative to pre-existing annuity income,
higher levels of risk aversion, greater investment volatility $\sigma$ and
better health status (i.e. lower subjective mortality rates) will all serve to
increase the amount that is voluntarily annuitized. And, although the
magnitude of these results depend on the specific parameters involved, all of
these comparative statics are consistent with our qualitative intuition and
should therefore re-enforce normative investment advice.
\end{itemize}

\subsection{Directions for Future Research}

That our paper raises a number of questions that should be considered in any
future research on the topic, which we will now elaborate on.

First, by far the strongest assumption we have made in our modeling for both
the restricted all-or-nothing market and the anything-anytime environment, is
that the risk-free rate and the market risk premium are assumed constant.
Thus, we have abstracted from any term structure effects, or predictability in
the evolution of interest rates as well as stochasticity of investment
volatility $\sigma$ and the market risk premium. Yet, recent models of asset
allocation and portfolio choice have gone well beyond the classic Mertonian
framework, which was the foundation of our analysis. The next step, from our
perspective, is to enhance the financial chassis of our model and allow for
more complicated market dynamics.

Indeed, a number of practitioners have advocated that people refrain from
annuitizing when interest rates are low and that annuitization is more
appealing when interest rates are high. Other commentators have advocated a
dollar-cost averaging approach to annuitization to smooth out the interest
rate risk, which resonates with the results from our anything-anytime
analysis. Obviously, one might question the precise definition of high and low
interest rates in this context, but it would certainly be interesting to
investigate the impact of assuming a mean-reverting process for interest rates
and possibly the yield curve as a whole. This would necessitate modeling the
evolution of nominal versus real interest rates as well as the behavior of
inflation. We, thus, envision an advanced model in which real
(inflation-adjusted) and nominal life annuities co-exist in the optimal
portfolio. This complicates our basic model by introducing at least one more
state variable in the PDE/ODE, which is why we leave for future research.

Likewise, a recent innovation in the U.S. retirement income market is the
introduction of guarantee living benefits, which are essentially staggered put
options on a portfolio that promise a minimal level of income for as long the
annuitant lives. These products which fall under the industry label of
Guaranteed Minimum Withdrawal Benefits (GMWB), have some longevity insurance
features and some derivative securities features. These products which are
part of the trillion dollar Variable Annuity (VA) industry in the U.S. are
growing in popularity and might actually compete with conventional life
annuities as a way of generating a sustainable retirement income. Further
research would examine the optimal demand and asset allocation including these
hybrid GMWB products.

Furthermore, given the normative focus on this paper, we have ignored the
positive equilibrium implications. The main question in this case would be how
annuity prices would be affected by individuals' desire to annuitize at an
optimal time. Indeed, the payoff from conventional financial assets are not
age- or mortality-dependent and thus do not depend on the demographic
structure or health status of the marginal investor. Under the discretionary
and voluntary life annuities that we have analyzed, one can envision a
situation in which very few people purchase a life annuity at age 50, which
thus reduces the pool of individuals across which mortality risk can be
diversified via the law of large numbers. This will have immediate implication
on the pricing curve and the objective hazard rate, which we have taken as
given. A detailed equilibrium analysis would attempt to derive the market's
objective hazard as a function of the heterogeneous mix of participant's
subjective hazard rates. But, this is far beyond the scope of the current paper.

On a related note, our model implicitly assumes that mortality rates are fully
predictable in the future and that we are able to specify a survival function
and annuity pricing equation during the entire horizon, conditional on the
value of interest rates. In other words, we assume that the objective hazard
rate is deterministic. However, there is a growing body of empirical and
theoretical literature that argues that mortality risk is priced in
equilibrium. In the extreme, this would imply that if one delays annuitization
one runs the risk that annuity prices will actually increase, even though the
individual has aged. This, of course, would introduce yet another variable in
the decision and further complicate the analysis of the optimal age at which
to annuitize. That said, with the impending retirement of American baby
boomers and the industrial shift from Defined Benefit (DB) to Defined
Contribution (DC) pension plans, we believe these issues will demand further
academic attention as they take on greater practical importance.%

%TCIMACRO{\TeXButton{TeX field}{\newpage}}%
%BeginExpansion
\newpage
%EndExpansion

\section{Appendix}

\subsection{Appendix A: Impact of Objective vs. Subjective Health}

In this appendix, we show that if the subjective force of mortality varies
slightly from the objective force to the extent that $\bar{a}_{x}^{S}<2\bar
{a}_{x}^{O}$ for all $x$, then the optimal time of annuitization increases
from the $T$ given by the zero of the right-hand side of (\ref{3.7}). In
particular, if the individual is less healthy than normal ($\lambda_{x}%
^{S}>\lambda_{x}^{O}$ for all $x$), then $\bar{a}_{x}^{S}<\bar{a}_{x}^{O}$ for
all $x$, from which it follows that the optimal time of annuitization is
delayed. Also, if the individual is more healthy than normal but only to the
extent that $\bar{a}_{x}^{S}<2\bar{a}_{x}^{O}$ for all $x$, then the optimal
time of annuitization is delayed.

Suppose $\bar{a}_{x+T}^{S}=\bar{a}_{x+T}^{O}+\varepsilon$ for some small
$\varepsilon$, not necessarily positive. Then, equation (\ref{3.7}) at the
critical value $T$ becomes%

\begin{equation}
0=\left[  \frac{\gamma}{1-\gamma}\left(  \frac{\bar{a}_{x+T}^{O}+\varepsilon
}{\bar{a}_{x+T}^{O}}\right)  ^{-\frac{1-\gamma}{\gamma}}-\frac{1}{1-\gamma
}+\frac{\bar{a}_{x+T}^{O}+\varepsilon}{\bar{a}_{x+T}^{O}}\right]  +\left(
\bar{a}_{x+T}^{O}+\varepsilon\right)  \left[  \delta-\left(  r+\lambda
_{x+T}^{O}\right)  \right]  .
\end{equation}

\noindent We can simplify this equation to%

\begin{align}
0  &  =\left(  \bar{a}_{x+T}^{O}+\varepsilon\right)  \left[  \delta-\left(
r+\lambda_{x+T}^{O}\right)  \right]  -\frac{1}{2}\left(  -\frac{1-\gamma
}{\gamma}-1\right)  \left(  \frac{\varepsilon}{\bar{a}_{x+T}^{O}}\right)
^{2}\nonumber\\
&  -\frac{1}{6}\left(  -\frac{1-\gamma}{\gamma}-1\right)  \left(
-\frac{1-\gamma}{\gamma}-2\right)  \left(  \frac{\varepsilon}{\bar{a}%
_{x+T}^{O}}\right)  ^{3}+\ldots,
\end{align}

\noindent if $\frac{\varepsilon}{\bar{a}_{x+T}^{O}}$ lies between $-1$ and
$1$. Thus, by the mean value theorem, there exists $\varepsilon^{*}$ between
$0$ and $\frac{\varepsilon}{\bar{a}_{x+T}^{O}}$ such that%

\[
0=\left(  \bar{a}_{x+T}^{O}+\varepsilon\right)  \left[  \delta-\left(
r+\lambda_{x+T}^{O}\right)  \right]  +\frac{1}{2\gamma}\left(  \varepsilon
^{*}\right)  ^{2},
\]

\noindent or equivalently,%

\begin{equation}
0=\left[  \delta-\left(  r+\lambda_{x+T}^{O}\right)  \right]  +\frac{\left(
\varepsilon^{\ast}\right)  ^{2}}{2\gamma}\frac{1}{\bar{a}_{x+T}^{O}%
(1+\frac{\varepsilon}{\bar{a}_{x+T}^{O}})}.
\end{equation}

\noindent The second term of the above equation is positive (and small)
regardless of the sign of $\varepsilon$. Thus, $T$ is determined by setting
$\left[  \delta-\left(  r+\lambda_{x+T}^{O}\right)  \right]  $ equal to a
negative number. It follows that, for $\lambda_{x}^{O}$ increasing with
respect to $x$, this value of $T$ will be larger than the zero of (\ref{3.8}).

Note that a sufficient condition for the above result is that $\frac
{\varepsilon}{\bar{a}_{x+T}^{O}}$ lie between $-1$ and $1$. Without
difficulty, one can show that this requirement is equivalent to $\bar{a}%
_{x+T}^{S}<2\bar{a}_{x+T}^{O}$. For less healthy people ($\lambda_{x}%
^{S}>\lambda_{x}^{O}$ for all $x$), we have $\bar{a}_{x}^{S}<\bar{a}_{x}^{O}$
for all x, so $\bar{a}_{x+T}^{S}<2\bar{a}_{x+T}^{O}$ holds automatically.
Also, there is some leeway in this inequality, so that even healthier people
might have that $\bar{a}_{x+T}^{S}<2\bar{a}_{x+T}^{O}$. Even when this
inequality does not hold, we conjecture that we still have a delay in the time
of annuitization beyond that given by the zero of the right-hand side of
(\ref{3.8}), as we see in one of the examples in Section 5.

\subsection{Appendix B: Probability of Reduced Payments}

The probability that consumption (as a percentage of initial wealth) at
optimal time of annuitization is $p$\% less than the consumption if one
annuitizes immediately equals%

\begin{align}
&  \mathbf{P}\left(  \frac{W_{T}^{\ast}}{\bar{a}_{x+T}^{O}}<\left(
1-0.01p\right)  \frac{w}{\bar{a}_{x}^{O}}\mid W_{0}=w\right) \nonumber\\
&  =\mathbf{P}\left(  we^{\left(  2\delta-r\frac{\left(  \mu-r\right)  ^{2}%
}{2\sigma^{2\gamma^{2}}}\right)  T-\int_{0}^{T}k(s)ds+\frac{\mu-r}%
{\sigma\gamma}B_{T}}<\left(  1-0.01p\right)  w\frac{\bar{a}_{x+T}^{O}}{\bar
{a}_{x}^{O}}\right) \nonumber\\
&  =\mathbf{P}\left(  e^{\frac{^{\mu-r}}{\sigma\gamma}B_{T}}<\frac{\left(
1-0.01p\right)  \bar{a}_{x+T}^{O}}{\bar{a}_{x}^{O}}e^{^{-\left(
2\delta-r-\frac{\left(  \mu-r\right)  ^{2}}{2\sigma^{2\gamma^{2}}}\right)
T+\int_{0}^{T}k(s)ds}}\right) \nonumber\\
&  =\mathbf{P}\left(  B_{T}<\frac{\ln\left(  \frac{\left(  1-0.01p\right)
\bar{a}_{x+T}^{O}}{\bar{a}_{x}^{O}}\right)  -\left(  2\delta-r-\frac{\left(
\mu-r\right)  ^{2}}{2\sigma^{2\gamma^{2}}}\right)  T+\int_{0}^{T}k(s)ds}%
{\frac{\mu-r}{\sigma\gamma}}\right) \nonumber\\
&  =\Phi\left(  \frac{\ln\left(  \frac{\left(  1-0.01p\right)  \bar{a}%
_{x+T}^{O}}{\bar{a}_{x}^{O}}\right)  -\left(  2\delta-r-\frac{\left(
\mu-r\right)  ^{2}}{2\sigma^{2\gamma^{2}}}\right)  T+\int_{0}^{T}k(s)ds}%
{\frac{\mu-r}{\sigma\gamma}\sqrt{T}}\right)  .
\end{align}

\noindent Here, $\Phi$ denotes the cumulative distribution function of the
standard normal.

\subsection{Appendix C: Variable and Fixed Annuity Benefits}

In the body of the paper, we assumed that the only life annuities available
upon annuitization provided a fixed payout. In this appendix, we examine a
market in which annuity `wrappers' are available on all asset classes with
full asset allocation mobility. We investigate the optimal consumption,
investment, and annuitization policies in this market. We introduce the symbol
$\beta$ to represent the proportion of wealth at the time of annuitization
that is invested in the variable annuity, so that $1-\beta$ is the proportion
invested in the fixed annuity. We assume that the mix between the variable and
fixed annuities, namely $\beta$ versus $1-\beta$, stays fixed throughout the
remaining life of the annuitant, which is a so-called \textit{money mix} plan
and has certain optimality features as shown by Charupat and Milevsky (2002).
Again, we consider CRRA utility and provide formulas for the power utility
case. One can easily deal with the logarithmic case by letting $\lambda$
approach 1 in the consumption, investment, and annuitization policies under
power utility $u(c)=\frac{1}{1-\gamma} c^{1-\gamma},\gamma> 0, \gamma\ne1.$

To further capture the salient features of this product, we assume that the
provider of the annuity has a nonzero insurance load on the fixed annuity such
that the effective ``return'' on the fixed annuity is $r^{\prime}$, with
$r^{\prime}\leq r$. Similarly, there is a nonzero insurance load on the
variable annuity such that the drift on the variable annuity is $\mu^{\prime}$
with $\mu^{\prime}\leq\mu$ and $\mu^{\prime}-r^{\prime}\leq\mu-r$.

For the mixture of a variable and a fixed annuity, define the value function
$V$ by%

\begin{align}
V(w,t;T)  &  =\sup_{\left\{  c_{s,}\pi_{s,}\beta\right\}  } \mathbf{E}^{w, t}
\left[  \int_{t}^{T}e^{-r\left(  s-t\right)  } \, _{s-t}p_{x+t}^{S} \,
\frac{c_{s}^{1-\gamma}}{1-\gamma} \, ds\right. \label{3.11}\\
&  \quad+\left.  \int_{T}^{\infty}e^{-r\left(  s-t\right)  } \, _{s-t}%
p_{x+t}^{S} \, \frac{1}{1-\gamma}\left(  \frac{W_{T}}{\bar{a}_{x+T}%
^{O,r^{\prime}}} \, e^{\beta\left(  \mu^{\prime}-r^{\prime}-\frac{\beta
\sigma^{2}}{2}\right)  \left(  s-T\right)  +\beta\sigma\left(  B_{s}%
-B_{T}\right)  }\right)  ^{1-\gamma}ds ,\right] \nonumber
\end{align}

\noindent in which the second superscript on $\bar{a}_{x+T}^{O,r^{\prime}}$,
namely $r^{\prime}$, is the rate of discount used to calculate the actuarial
present value of the annuity.

We can deal with the choice of $\beta$ by noting that (Bj\"ork, 1998)%

\begin{align}
&  \mathbf{E}^{w, t} \left[  W_{T}^{1-\gamma}e^{\beta\left(  1-\gamma\right)
\left(  \mu^{\prime}-r^{\prime}-\frac{\beta\sigma^{2}}{2}\right)  \left(
s-T\right)  +\beta\left(  1-\gamma\right)  \sigma\left(  B_{s}-B_{T}\right)
}\right] \\
&  =\mathbf{E}^{w, t} \left[  W_{T}^{1-\gamma}\right]  e^{\beta\left(
1-\gamma\right)  \left(  \mu^{\prime}-r^{\prime}-\frac{\beta\gamma\sigma^{2}%
}{2}\right)  \left(  s-T\right)  . }\nonumber
\end{align}

\noindent Thus, the expectation is maximized if we maximize $\beta\left(
\mu^{\prime}-r^{\prime}-\frac{\beta\gamma\sigma^{2}}{2}\right)  $. It follows
that the optimal value of $\beta$ equals%

\begin{equation}
\beta^{\ast}=\frac{\mu^{\prime}-r^{\prime}}{\sigma^{2}}.
\end{equation}

\noindent Note that the optimal choice of $\beta$ is independent of the
optimal time $T$ of annuitization. Naturally, if $\beta^{*}$ is greater than
one, the holdings are truncated by the seller of the annuity at 100\% of the
risky stock.

It follows that $V$ solves the HJB equation%

\begin{equation}
\left\{
\begin{array}
[c]{l}%
\left(  r+\lambda_{x+t}^{S}\right)  V=V_{t}+\max\limits_{\pi}\left[  \frac
{1}{2}\sigma^{2}p^{2}V_{ww}+\left(  \mu-r\right)  \pi V_{w}\right]  +rwV_{w}
+\max\limits_{c\geq0}\left[  -cV_{w}+\frac{1}{1-\gamma}c^{1-\gamma}\right]
,\\
V\left(  w,T;T\right)  =\frac{1}{1-\gamma}\left(  \frac{w}{\bar{a}%
_{x+T}^{O,r^{\prime}}}\right)  ^{1-\gamma} \bar{a}_{x+T}^{S,r-\frac{\left(
1-\gamma\right)  \left(  \mu^{\prime}-r^{\prime}\right)  ^{2}}{s\sigma^{2}y}}.
\end{array}
\right.  \label{3.12}%
\end{equation}

\noindent Note that this is the same as the previous HJB equation, except that
the boundary condition reflects the mixture of the variable and fixed
annuities. The second superscript on the actuarial present value denotes the
rate at which the annuity payments are discounted. It follows that $V$ has the
form as that given in equation (\ref{3.4}), except that $\bar{a}_{x+T}%
^{S}=\bar{a}_{x+T}^{S,r}$ is replaced with
\begin{equation}
\bar{a}_{x+T}^{S,r-\frac{\left(  1-\gamma\right)  \left(  \mu^{\prime
}-r^{\prime}\right)  ^{2}}{s\sigma^{2}y}},
\end{equation}
and $\bar{a}_{x+T}^{O}=\bar{a}_{x+T}^{O,r}$ is replaced with $\bar{a}%
_{x+T}^{O,r^{\prime}}$. Also, note that the optimal investment policy is
similar in form to the one given in the body of the paper for the fixed-only
payout case. If there are no loads on the fixed and variable annuities, that
is, if $r^{\prime}=r$ and $\mu^{\prime}=\mu$, then the proportion of wealth
invested in the risky asset from before annuitization equals the proportion
after annuitization; however, in this case, the optimal strategy of the
individual is to annuitize immediately. This latter result follows from the
work of Yaari (1965). For a CRRA investor (with no bequest motives) with no
insurance loads on the annuities, the optimal mixture between risky and
risk-free assets is invariant to whether the portfolio is annuitized or not.

The derivative of $V$ with respect to $T$ is proportional to%

\begin{align}
\frac{\delta V}{\delta T}  &  \propto\left[  \frac{\gamma}{1-\gamma}\left(
\frac{\bar{a}_{x+T}^{S,r-\frac{\left(  1-\gamma\right)  \left(  \mu^{\prime
}-r^{\prime}\right)  ^{2}}{s\sigma^{2}y}}}{\bar{a}_{x+T}^{O,r^{\prime}}%
}\right)  ^{\frac{1-\gamma}{\gamma}}-\frac{1}{1-\gamma}+\frac{\bar{a}%
_{x+T}^{S,r-\frac{\left(  1-\gamma\right)  \left(  \mu^{\prime}-r^{\prime
}\right)  ^{2}}{s\sigma^{2}y}}}{\bar{a}_{x+T}^{O,r^{\prime}}}\right]
\label{3.13}\\
&  \qquad+\bar{a}_{x+T}^{S,r-\frac{\left(  1-\gamma\right)  \left(
\mu^{\prime}-r^{\prime}\right)  ^{2}}{s\sigma^{2}y}}\left[  \delta
-\delta^{\prime}-\lambda_{x+T}^{O}\right]  ,\nonumber
\end{align}

\noindent in which $\delta^{\prime}=r^{\prime}+\frac{\left(  \mu^{\prime
}-r^{\prime}\right)  ^{2}}{2\sigma^{2}\gamma}$. We can use this equation to
determine the optimal value of $T$ in any given situation. In Table 5, we
compare the optimal ages of annuitization and the imputed value of delaying
when the individual can only buy a fixed annuity (compare with Table 1) and
when the individual can buy a money mix of variable and fixed annuities.%

\[%
\begin{tabular}
[c]{||c||}\hline\hline
\textbf{Table 5a about here.}\\\hline\hline
\end{tabular}
\]
\hfill

We assume that the financial market and mortality are as described in Section
5, except that for the variable annuity, the insurer has a 100-basis-point
Mortality and Expense Risk Charge load on the return, so that the modified
drift is $\mu^{\prime}=0.11$, and for the fixed annuity, the insurer has a
50-basis-point spread on the return, so that the modified rate of return is
$r^{\prime}=0.055$.%

\[%
\begin{tabular}
[c]{||c||}\hline\hline
\textbf{Table 5b about here.}\\\hline\hline
\end{tabular}
\]

Assume that the individual has a CRRA of $\gamma=2$, from which it follows
that the individual will invest 75.0\% in the risky stock before annuitization
and 68.7\% in the variable annuity after annuitization.

\subsection{Appendix D: Escalating Annuity Benefit}

Suppose that the individual can buy an escalating annuity. An escalating
annuity is one for which the payments increase at a (constant) rate $g$. These
are known as COLA (Cost Of Living Adjustment) annuities and are available from
vendors that sell fixed annuities. These escalating annuities are popular as a
hedge against (expected) inflation, since it is virtually impossible to obtain
true inflation-linked annuities in the U.S. The actuarial present value of an
escalating annuity can be written $\bar{a}_{x}^{r-g};$ that is, the rate of
discount $r$ is reduced by the rate of increase of the payments $g$. As in the
previous two sections, we consider CRRA utility and provide formulas for the
power utility case. Thus, we can define the corresponding value function by%

\begin{align}
V\left(  w,t;T\right)   &  =\sup_{\left\{  c_{s},\pi_{s},g\right\}  }\left[
E\int_{t}^{T}e^{-r\left(  s-t\right)  }\left.  _{s-t}\right.  p_{x+t}^{S}%
\frac{1}{1-\gamma}c_{s}^{1-\gamma}ds\right. \label{3.14}\\
&  \left.  +\left.  \int_{T}^{\infty}e^{-r\left(  s-t\right)  }\left.
_{s-t}\right.  p_{x+t}^{S}\frac{1}{1-\gamma}\left(  \frac{W_{T}}{\bar{a}%
_{x+T}^{O,r-g}}e^{g\left(  s-T\right)  }\right)  ^{1-\gamma}ds\right|
W_{t}=w\right] \nonumber
\end{align}

\noindent This expression is maximized with respect to $g$ if%

\begin{equation}
\frac{1}{1-\gamma}\frac{\int_{0}^{\infty}e^{-\left(  r-\left(  1-\gamma
\right)  g\right)  s}}{\left[  \int_{0}^{\infty}e^{-\left(  r-g\right)
s}\left.  _{s}p_{x+T}^{O}ds\right.  \right]  ^{1-\gamma}}%
\end{equation}

\noindent is maximized. The derivative of this expression with respect to $g$
is proportional to%

\begin{equation}
\frac{\int_{0}^{\infty}se^{-\left(  r-\left(  1-\gamma\right)  g\right)
s}\left.  _{s}p_{x+T}^{S}ds\right.  }{\int_{0}^{\infty}e^{-\left(  r-\left(
1-\gamma\right)  g\right)  s}\left.  _{s}p_{x+T}^{S}ds\right.  }-\frac
{\int_{0}^{\infty}se^{-\left(  r-g\right)  s}\left.  _{s}p_{x+T}^{O}ds\right.
}{\int_{0}^{\infty}e^{-\left(  r-g\right)  s}\left.  _{s}p_{x+T}^{O}ds\right.
}.
\end{equation}

\noindent Note that this is a difference of expectations of \textquotedblleft%
$s$\textquotedblright\ with respect to two probability distributions. Also,
note that if $\lambda_{x}^{S}=\lambda_{x}^{O}-c$ for all $x$ and for some
constant $c$, then the optimal value of $g$ is $g^{\ast}=\frac{c}{y}$. For
example, if the individual is healthier to the extent that the subjective
hazard rate is 0.01 less than the objective (pricing) hazard rate, then
optimal rate of increase of the annuity payments (once the individual
annuitizes his or her wealth) is $\frac{0.01}{\gamma}$. Note that in general,
\textit{healthier} people will want to buy escalating annuities with a
\textit{positive} $g$, while \textit{sicker }people will want to buy
escalating annuities with a \textit{negative} $g$. This makes sense because
healthier people anticipate living longer than normal, so they will be able to
enjoy those larger annuity payments in the future. On the other hand, less
healthy people will not live as long, so they demand higher payments now.

Table 6 provides a numerical example for an individual who believes he or she
is healthier than normal with $f=-0.5$; that is, the person has one-half the
mortality rate of the average person. Suppose that the CRRA is $\gamma=1.5$.
We compare these numbers with those when the individual can buy only a fixed
annuity. It turns out that the optimal rate of escalation $g=2\%$ (to two
decimal places) for all ages and for both genders.%
\[%
\begin{tabular}
[c]{||c||}\hline\hline
\textbf{Table 6 about here.}\\\hline\hline
\end{tabular}
\]

\noindent Note that the individual is willing to annuitize earlier if there is
a 2\% escalating annuity available; however, there is still an advantage to wait.%

%TCIMACRO{\TeXButton{TeX field}{\newpage}}%
%BeginExpansion
\newpage
%EndExpansion

\begin{center}
\textbf{Table 1a: Optimal Age to Annuitize and Value of Delay under Low Risk
Aversion}

\vspace{0.25in}
\begin{tabular}
[c]{|l|l|l|l|l|l|l|}\hline
& \multicolumn{3}{|l|}{$\gamma=1$, FEMALE (MALE)} &
\multicolumn{3}{|l|}{$\gamma=2$, FEMALE (MALE)}\\\hline
\textbf{Age} & \textbf{Optimal} & \textbf{Value} & \textbf{Prob. Lower} &
\textbf{Optimal} & \textbf{Value} & \textbf{Prob. Lower}\\
& \textbf{Age} & \textbf{of Delay} & \textbf{Annuity} & \textbf{Age} &
\textbf{of Delay} & \textbf{Annuity}\\\hline
\textbf{60} & 84.5 (80.3) & 44.0 (32.0)\% & 0.311 (0.353) & 78.4 (73.0) & 15.3
(8.9)\% & 0.268 (0.321)\\\hline
\textbf{65} & 84.5 (80.3) & 33.4 (21.9) & 0.346 (0.391) & 78.4 (73.0) & 10.3
(4.3) & 0.310 (0.372)\\\hline
\textbf{70} & 84.5 (80.3) & 22.7 (12.3) & 0.385 (0.431) & 78.4 (73.0) & 5.2
(0.8) & 0.362 (0.435)\\\hline
\textbf{75} & 84.5 (80.3) & 12.3 (4.2) & 0.429 (0.470) & 78.4 (Now) & 1.2
(0.0) & 0.428 (N/a)\\\hline
\textbf{80} & 84.5 (80.3) & 3.7 (0.02) & 0.473 (0.500) & Now (Now) & neg.
(neg.) & N/a (N/a)\\\hline
\textbf{85} & Now (Now) & neg. (neg.) & N/a (N/a) & Now (Now) & neg. (neg.) &
N/a (N/a)\\\hline
\end{tabular}
$\mathbf{\ }$
\end{center}

Notes: All-or-Nothing Market: The value of the option to delay annuitization
for males and females with a coefficient of relative risk aversion of $\gamma=
1$ and $\gamma= 2$. We assume the non-annuitized funds are invested in a risky
asset with drift $\mu= 0.12$ and volatility $\sigma= 0.20$. The risk-free rate
is $r = 0.06$. The mortality is assumed to be Gompertz-Makeham fit to the
IAM2000 table with projection Scale G. For example, a 70-year-old female with
a coefficient of relative risk aversion of $\gamma= 2$, will effectively
suffer a utility-equivalent 5.2\% loss of her wealth if she chooses to
annuitize immediately. The optimal time for her to annuitize is at age 78.4.
The table also gives the probability of deferral failure, namely the
probability that the annuity purchased at the optimal age will provide less
income that an annuity bought right now.%

%TCIMACRO{\TeXButton{TeX field}{\clearpage}}%
%BeginExpansion
\clearpage
%EndExpansion

\begin{center}
\textbf{Table 1b: Optimal Age to Annuitize and Value of Delay under Higher
Risk Aversion}%

\begin{tabular}
[c]{|c|c|c|c|c|}\hline
& \multicolumn{4}{|c|}{Coefficient of Relative Risk Aversion $\gamma=5$%
}\\\hline
& \multicolumn{2}{|c|}{{MALE}} & \multicolumn{2}{|c|}{{FEMALE}}\\\hline
\textbf{Current Age} & \textbf{Optimal Age} & \textbf{Value of Delay} &
\textbf{Optimal Age} & \textbf{Value of Delay}\\\hline
\textbf{60} & 63.4 & 0.41\% & 70.4 & 2.94\%\\\hline
\textbf{65} & Now & neg. & 70.4 & 1.04\%\\\hline
\textbf{70} & Now & neg. & 70.4 & 0.01\%\\\hline
\textbf{75} & Now & neg. & Now & neg.\\\hline
\end{tabular}

\end{center}

Notes: All-or-nothing market. The risk-free rate is $r=0.06$, the drift is
$\mu=0.12$, and the volatility is $\sigma=0.20$. We use Gompertz-Makeham
mortality with $m=88.15$, $b=10.5$ for males and $m=92.63$, $b=8.78$ for
females. Notice that even at higher levels of risk aversion, males (and
especially females) do not annuitize prior to age 60.%

%TCIMACRO{\TeXButton{TeX field}{\clearpage}}%
%BeginExpansion
\clearpage
%EndExpansion

\begin{center}
\textbf{Table 2: Assuming Optimal Behavior, What is the Probability of Higher
Consumption Upon Delay?}%

\begin{tabular}
[c]{|l|l|l|}\hline
& $\gamma=1,$ FEMALE (MALE) & $\gamma=2$, FEMALE (MALE)\\\hline
\textbf{Age} & \textbf{Prob. of consuming at least} & \textbf{Prob. of
consuming at least}\\
& \textbf{20\% more than initial annuity} & \textbf{20\% more than initial
annuity}\\\hline
\textbf{60} & 0.644 (0.596) & 0.631 (0.551)\\\hline
\textbf{65} & 0.602 (0.549) & 0.565 (0.459)\\\hline
\textbf{70} & 0.552 (0.494) & 0.474 (0.296)\\\hline
\textbf{75} & 0.493 (0.425) & 0.316 (0.133)\\\hline
\textbf{80} & 0.414 (0.137) & N/a (N/a)\\\hline
\textbf{85} & N/a (N/a) & N/a (N/a)\\\hline
\end{tabular}

\end{center}

Notes: All-or-nothing market. Assuming the individual self-annuitizes and
defers the purchase of a life annuity to the optimal age, this table indicates
the probability of consuming at least 20\% more at the time of annuitization,
compared to if one annuitizes immediately. Thus, for example, a female (male)
at age 65 with a coefficient of relative risk aversion of =1 has a 64.4\%
(59.6\%) chance of creating a 20\% larger annuity flow.%

%TCIMACRO{\TeXButton{TeX field}{\clearpage}}%
%BeginExpansion
\clearpage
%EndExpansion

\begin{center}
\textbf{Table 3: How Does Subjective Health Status Impact Optimal Behavior?}%

\begin{tabular}
[c]{|l|l|l|l|l|}\hline
$f$ & \textbf{Optimal Age of} & \textbf{Value} & \textbf{Consumption Rate} &
\textbf{Consumption Rate}\\
More/Less Healthy & \textbf{annuitization} & \textbf{of Delay} &
\textbf{Before Annuitization} & \textbf{After Annuitization}\\\hline
-1.0 & 78.28 & 13.79\% & 7.55\% & 13.38\%\\\hline
-0.8 & 74.58 & 10.54 & 7.95 & 11.79\\\hline
-0.6 & 73.71 & 9.68 & 8.18 & 11.47\\\hline
-0.4 & 73.29 & 9.23 & 8.37 & 11.33\\\hline
-0.2 & 73.09 & 8.99 & 8.54 & 11.26\\\hline
0.0 & 73.03 & 8.87 & 8.70 & 11.24\\\hline
0.2 & 73.08 & 8.84 & 8.85 & 11.26\\\hline
0.5 & 73.31 & 8.93 & 9.06 & 11.33\\\hline
1.0 & 74.04 & 9.34 & 9.38 & 11.59\\\hline
1.5 & 75.21 & 10.00 & 9.68 & 12.03\\\hline
2.0 & 76.96 & 10.89 & 9.98 & 12.76\\\hline
2.5 & 79.71 & 12.01 & 10.26 & 14.12\\\hline
3.0 & 85.38 & 13.38 & 10.55 & 18.01\\\hline
\end{tabular}

\end{center}

Notes: All-or-Nothing Market: The imputed value of delaying annuitization for
a male, aged 60 with CRRA of $\gamma=2$. We assume the funds are invested in a
risky asset with drift $\mu=0.12$ and volatility $\sigma=0.20$. The risk-free
rate is $r=0.06$. The mortality is assumed to be Gompertz-Makeham fit to the
IAM2000 table with projection Scale G. Thus, for example, if the individual's
subjective hazard rate is 20\% higher (i.e. less healthy) than the mortality
table used by the insurance company to price annuities, the optimal
annuitization point is at age 73.1, and the value of the option is 8.84\% of
his wealth at age 60. While the 60-year-old male waits to annuitize, he
consumes at the rate of 8.85\% of assets, and once he purchases the fixed
annuity, his consumption rate - and standard of living - will increase to
11.26\% of assets.%

%TCIMACRO{\TeXButton{TeX field}{\clearpage}}%
%BeginExpansion
\clearpage
%EndExpansion

\begin{center}
\textbf{Table 4a: How Does Wealth and Risk Aversion Affect Annuitization?}%

\begin{tabular}
[c]{|c|c|c|c|c|c|}\hline
\multicolumn{6}{|c|}{Amount of Money Spent on Annuities for Various Levels}\\
\multicolumn{6}{|c|}{of Wealth and Risk Aversion ($A=$ \$25,000)}\\\hline
& $\gamma=1.5$ & $\gamma=2.0$ & $\gamma=2.5$ & $\gamma=3.0$ & $\gamma
=5.0$\\\hline
Wealth & $\left(  z_{0}=3.273\right)  $ & $\left(  z_{0}=2.354\right)  $ &
$\left(  z_{0}=1.837\right)  $ & $\left(  z_{0}=1.506\right)  $ &
$(z_{0}=0.874)$\\\hline
\$1,000,000 & \$727,620 & \$792,020 & \$831,852 & \$858,901 &
\$914,176\\\hline
\$500,000 & \$331,384 & \$371,251 & \$395,909 & \$412,653 & \$446,871\\\hline
\$250,000 & \$133,266 & \$160,866 & \$177,937 & \$189,529 & \$213,218\\\hline
\$100,000 & \$14,395 & \$34,635 & \$47,154 & \$55,655 & \$73,027\\\hline
\$50,000 & \$0 & \$0 & \$3559 & \$11,030 & \$26,296\\\hline
\end{tabular}

\end{center}

Notes: Anything Anytime Market: Table 4 illustrates for various levels of
relative risk aversion $\gamma$, the critical value of the ratio of liquid
wealth to pre-existing annuity income $z_{0}=w/A$ above which the individual
will spend a lump sum to increase her annuity income. We also include the
amount the individual will spend on annuities, namely $(w-z_{0}%
A)/(1+(r+\lambda^{O})z_{0})$. We assume the following parameter values: The
force of mortality $\lambda^{S} = \lambda^{O}=0.04$, which implies a life
expectancy of 25 years, the riskless rate of return is $r=0.04$, the risky
asset's drift is $\mu=0.08$, and the risky asset's volatility is $\sigma
=0.20$. Note that in contrast to the restricted all-or-nothing market, the
individual immediately annuitizes a base level of income and then gradually
annuitizes more as his wealth breaches higher levels.%

%TCIMACRO{\TeXButton{TeX field}{\clearpage}}%
%BeginExpansion
\clearpage
%EndExpansion

\begin{center}
\textbf{Table 4b: How Does Wealth and Risk Aversion Affect Annuitization?}

\vspace{0.25in}
\begin{tabular}
[c]{|c|c|c|c|c|c|}\hline
\multicolumn{6}{|c|}{Amount of Money Spent on Annuities for Various Levels}\\
\multicolumn{6}{|c|}{of Wealth and Risk Aversion ($A=$ \$50,000)}\\\hline
& $\gamma=1.5$ & $\gamma=2.0$ & $\gamma=2.5$ & $\gamma=3.0$ & $\gamma
=5.0$\\\hline
Wealth & $\left(  z_{0}=3.273\right)  $ & $\left(  z_{0}=2.354\right)  $ &
$\left(  z_{0}=1.837\right)  $ & $\left(  z_{0}=1.506\right)  $ &
$(z_{0}=0.874)$\\\hline
\$1,000,000 & \$662,802 & \$742,477 & \$791,789 & \$825,271 &
\$893,741\\\hline
\$500,000 & \$266,555 & \$321,715 & \$355,854 & \$379,034 & \$426,436\\\hline
\$250,000 & \$68,432 & \$111,334 & \$137,886 & \$155,915 & \$192,784\\\hline
\$100,000 & \$0 & \$0 & \$7,106 & \$22,044 & \$52,592\\\hline
\$50,000 & \$0 & \$0 & \$0 & \$0 & \$5,862\\\hline
\end{tabular}

\end{center}

Notes: Anything anytime market. In Table 4a, the pre-existing annuity income
$A$ equals \$25,000, while in Table 4b, $A$ is doubled to \$50,000.
Intuitively, the greater the level of pre-existing annuity income the less
liquid wealth must be annuitized to provide the optimal annuitized consumption stream.%

%TCIMACRO{\TeXButton{TeX field}{\clearpage}}%
%BeginExpansion
\clearpage
%EndExpansion

\begin{center}
\textbf{Table 4c: How Does Investment Volatility Affect Annuitization?}%

\begin{tabular}
[c]{|c|c|c|}\hline
& \multicolumn{2}{|c|}{Wealth of \$1,000,000 and Initial Annuity Income of
\$40,000}\\\hline
\textbf{Investment Volatility} & Low Risk Aversion ($\gamma=2\,$) & High Risk
Aversion ($\gamma=5$)\\\hline
$\sigma=0.12$ & \$12,692 & \$496,789\\\hline
$\sigma=0.14$ & \$164,292 & \$598,755\\\hline
$\sigma=0.16$ & \$289,253 & \$672,235\\\hline
$\sigma=0.18$ & \$390,628 & \$726,853\\\hline
$\sigma=0.20$ & \$472,871 & \$768,568\\\hline
\end{tabular}

\end{center}

Notes: $r=0.05$ and $\mu=0.12$, with $\lambda^{O}=\lambda^{S}=0.04$. At higher
levels of investment volatility (risk), a greater amount is annuitized immediately.%

%TCIMACRO{\TeXButton{TeX field}{\clearpage}}%
%BeginExpansion
\clearpage
%EndExpansion

\begin{center}
\textbf{Table 4d: How Does Subjective Health Status Affect Annuitization?}%

\begin{tabular}
[c]{|c|c|c|}\hline
& \multicolumn{2}{|c|}{Wealth of \$1,000,000 and Initial Annuity Income of
\$40,000}\\\hline
\textbf{Subjective Hazard Rate} & Low Risk Aversion ($\gamma=2\,$) & High Risk
Aversion ($\gamma=5$)\\\hline
$\lambda^{S}=0.030$ & \$574,840 & \$817,383\\\hline
$\lambda^{S}=0.035$ & \$563,603 & \$812,222\\\hline
$\lambda^{S}=0.040$ & \$551,941 & \$806,842\\\hline
$\lambda^{S}=0.045$ & \$539,862 & \$801,242\\\hline
$\lambda^{S}=0.050$ & \$527,375 & \$795,423\\\hline
$\lambda^{S}=0.055$ & \$514,496 & \$789,388\\\hline
\end{tabular}

\end{center}

Notes: $r=0.05$, $\mu=0.10$, and $\sigma=0.16$, with $\lambda^{O}=0.04$. The
higher (less healthy) hazard rate leads to reduced levels of annuitization.%

%TCIMACRO{\TeXButton{TeX field}{\clearpage}}%
%BeginExpansion
\clearpage
%EndExpansion

\begin{center}
\textbf{Table 5a: All-or-Nothing Decision with Variable and Fixed Immediate
Annuities under Low Risk Aversion}%

\begin{tabular}
[c]{|l|l|l|l|l|}\hline
& \multicolumn{2}{|l|}{Fixed Annuity,} & \multicolumn{2}{|l|}{Mixture of
Variable (68.7\%) and}\\
& \multicolumn{2}{|l|}{FEMALE (MALE)} & \multicolumn{2}{|l|}{Fixed Annuities
(31.3\%),}\\
& \multicolumn{2}{|l|}{} & \multicolumn{2}{|l|}{FEMALE (MALE)}\\\hline
Age & Optimal age of & Value of & Optimal age of & Value of\\\hline
& annuitization & delay & Annuitization & delay\\\hline
60 & 80.2 (75.2) & 21.0 (13.4)\% & 70.8 (64.1) & 3.4 (0.6)\%\\\hline
65 & 80.2 (75.2) & 14.8 (7.5) & 70.8 (Now) & 1.3 (neg.)\\\hline
70 & 80.2 (75.2) & 8.5 (2.5) & 70.8 (Now) & 0.04 (neg.)\\\hline
75 & 0.2 (75.2) & 2.9 (0.003) & Now (Now) & neg. (neg.)\\\hline
\end{tabular}

\end{center}

Notes: In an all-or-nothing annuitization environment -- where both fixed and
variable annuities are available with complete asset allocation flexibility --
this table illustrates the imputed value of delaying annuitization for males
and females with a CRRA of $\gamma=2$. We assume that the non-annuitized funds
are invested in a risky asset with drift $\mu=0.12$ and volatility
$\sigma=0.20$. The risk-free rate is $r=0.06$. We introduce insurance loads on
the variable and fixed annuities on the order of 100 basis points and 50 basis
points, respectively. The mortality is assumed to be Gompertz fit to the
IAM2000 table with projection Scale G%

%TCIMACRO{\TeXButton{TeX field}{\clearpage}}%
%BeginExpansion
\clearpage
%EndExpansion

\begin{center}
\textbf{Table 5b: All-or-Nothing Decision under Various Levels of Insurance
Fees}%

\begin{tabular}
[c]{|c|c|c|c|c|}\hline
& \multicolumn{2}{|c|}{Low Risk Aversion ($\gamma=2$)} &
\multicolumn{2}{|c|}{High Risk Aversion ($\gamma=5$)}\\\hline
Insurance Fees Deducted & FEMALE & MALE & FEMALE & MALE\\\hline
50 b.p. & 67.2 & 60.0 & 66.9 & 60.0\\\hline
100 b.p. & 70.8 & 64.0 & 68.4 & 61.2\\\hline
125 b.p. & 72.1 & 65.6 & 69.1 & 61.9\\\hline
150 b.p. & 73.2 & 66.9 & 69.6 & 62.6\\\hline
200 b.p. & 74.9 & 69.0 & 70.6 & 63.8\\\hline
\end{tabular}

\end{center}

Notes: Assumes the same parameter values as Table 5a.%

%TCIMACRO{\TeXButton{TeX field}{\clearpage}}%
%BeginExpansion
\clearpage
%EndExpansion

\begin{center}
\textbf{Table 6: All-or-Nothing Decision with Variable and Fixed Immediate
Escalating Annuities}%

\begin{tabular}
[c]{|l|l|l|l|l|}\hline
& \multicolumn{2}{|l|}{Fixed Annuity} & \multicolumn{2}{|l|}{2\% Escalating
Annuity}\\
& \multicolumn{2}{|l|}{FEMALE (MALE)} & \multicolumn{2}{|l|}{FEMALE
(MALE)}\\\hline
Age & Optimal Age of & Value of & Optimal Age of & Value of\\
& Annuitization & Delay & Annuitization & Delay\\\hline
60 & 80.9 (76.1) & 23.68 (15.59)\% & 78.5 (73.2) & 17.41 (9.61)\%\\\hline
65 & 80.9 (76.1) & 17.05 (9.24) & 78.5 (73.2) & 11.50 (4.88)\\\hline
70 & 80.9 (76.1) & 10.22 (3.57) & 78.5 (73.2) & 5.80 (0.96)\\\hline
75 & 80.9 (76.1) & 3.96 (0.15) & 78.5 (Now) & 1.29 (0.00)\\\hline
\end{tabular}

\end{center}

Notes: In an all-or-nothing annuitization environment with escalating
annuities available, this table illustrates the value of delay for males and
females with a CRRA of $\gamma=1.5$. We assume the liquid funds are invested
in a risky asset with drift $\mu=0.12$ and volatility $\sigma=0.20$. The
risk-free rate is $r=0.06$. The mortality is Gompertz-Makeham fit to the
IAM2000 table with projection Scale G, while the individual has subjective
mortality beliefs equal to one-half of the objective mortality. Note that the
availability of an increasing annuity -- which better matches the desired
consumption profile -- accelerates the optimal age of annuitization and
reduces the option value to wait.%

%TCIMACRO{\TeXButton{TeX field}{\clearpage}}%
%BeginExpansion
\clearpage
%EndExpansion

\textbf{Figure \#1}%

%TCIMACRO{\FRAME{itbpFU}{437.75pt}{299.9375pt}{0pt}{\Qcb{{}}}{}{Figure1}%
%maintain-aspect-ratio TRUE;  display "USEDEF";  valid_file "T";
%width 437.75pt;  height 299.9375pt;  depth 0pt;  original-width 433.375pt;
%original-height 296.3125pt;  cropleft "0";  croptop "1";  cropright "1";
%cropbottom "0";  tempfilename 'J4OEWQ02.wmf';tempfile-properties "XPR";}}}%
%BeginExpansion
\raisebox{-0pt}{\parbox[b]{437.75pt}{\begin{center}
\includegraphics[
natheight=296.312500pt,
natwidth=433.375000pt,
height=299.9375pt,
width=437.75pt
]%
{J4OEWQ02.wmf}%
\\
{}%
\end{center}}}%
%EndExpansion

Notes: This figure shows the probability density function of the
future-lifetime random variable under an analytic Gompertz-Makeham hazard rate
that is fitted to the Individual Annuity Mortality Table 2000 with projection
scale G. For males, the `best fitting' parameters are $(m,b)=(88.18,10.5)$ and
for females they are $(92.63,8.78)$.%

%TCIMACRO{\TeXButton{TeX field}{\newpage}}%
%BeginExpansion
\newpage
%EndExpansion

\textbf{Figure \#2}%

%TCIMACRO{\FRAME{itbpFU}{437.75pt}{299.9375pt}{0pt}{\Qcb{{}}}{}{Figure
%2}{\special{ language "Scientific Word";  type "GRAPHIC";
%maintain-aspect-ratio TRUE;  display "USEDEF";  valid_file "T";
%width 437.75pt;  height 299.9375pt;  depth 0pt;  original-width 433.375pt;
%original-height 296.3125pt;  cropleft "0";  croptop "1";  cropright "1";
%cropbottom "0";  tempfilename 'J4OEWQ03.wmf';tempfile-properties "XPR";}}}%
%BeginExpansion
\raisebox{-0pt}{\parbox[b]{437.75pt}{\begin{center}
\includegraphics[
natheight=296.312500pt,
natwidth=433.375000pt,
height=299.9375pt,
width=437.75pt
]%
{J4OEWQ03.wmf}%
\\
{}%
\end{center}}}%
%EndExpansion

Notes: This figure shows the expected consumption of a 60-year-old male who
believes he is 20\% more healthy than average population rates; 8.34\% is the
rate of consumption if he annuitizes his wealth at age 60. We also display the
25th and 75th percentiles of the distribution of consumption between ages 60
and 75.%

%TCIMACRO{\TeXButton{TeX field}{\clearpage}}%
%BeginExpansion
\clearpage
%EndExpansion

\end{document}